\documentclass[twocolumn]{aastex63}

\usepackage{amsmath}
\let\tablenum\relax
\usepackage[separate-uncertainty,multi-part-units = single]{siunitx}
\usepackage{gensymb}

\shorttitle{Dynamical Mass of HD 984 B}
\shortauthors{Franson et al.}

\begin{document}

\title{Dynamical Mass of the Young Substellar Companion HD 984 B}

\correspondingauthor{Kyle Franson}
\email{kfranson@utexas.edu}

\author[0000-0003-4557-414X]{Kyle Franson}
\altaffiliation{NSF Graduate Research Fellow}
\affiliation{Department of Astronomy, The University of Texas at Austin, Austin, TX 78712, USA}

\author[0000-0003-2649-2288]{Brendan P. Bowler}
\affiliation{Department of Astronomy, The University of Texas at Austin, Austin, TX 78712, USA}

\author{Timothy D. Brandt}
\affiliation{Department of Physics, University of California, Santa Barbara, Santa Barbara, CA 93106, USA}

\author[0000-0001-9823-1445]{Trent J. Dupuy}
\affiliation{Institute for Astronomy, University of Edinburgh, Royal Observatory, Blackford Hill, Edinburgh, EH9 3HJ, UK}

\author[0000-0001-6532-6755]{Quang H. Tran}
\affiliation{Department of Astronomy, The University of Texas at Austin, Austin, TX 78712, USA}

\author[0000-0003-0168-3010]{G. Mirek Brandt}
\altaffiliation{NSF Graduate Research Fellow}
\affiliation{Department of Physics, University of California, Santa Barbara, Santa Barbara, CA 93106, USA}

\author[0000-0002-6845-9702]{Yiting Li}
\affiliation{Department of Physics, University of California, Santa Barbara, Santa Barbara, CA 93106, USA}

\author{Adam L. Kraus}
\affiliation{Department of Astronomy, The University of Texas at Austin, Austin, TX 78712, USA}

\begin{abstract}
Model-independent masses of substellar companions are critical tools to validate models of planet and brown dwarf cooling, test their input physics, and determine the formation and evolution of these objects. In this work, we measure the dynamical mass and orbit of the young substellar companion HD 984 B. We obtained new high-contrast imaging of the HD 984 system with Keck/NIRC2 which expands the baseline of relative astrometry from 3 to 8 years. We also present new radial velocities of the host star with the Habitable-Zone Planet Finder spectrograph at the Hobby-Eberly Telescope. Furthermore, HD 984 exhibits a significant proper motion difference between \emph{Hipparcos} and \emph{Gaia} EDR3. Our joint orbit fit of the relative astrometry, proper motions, and radial velocities yields a dynamical mass of $61 \pm 4$ $\mathrm{M_{Jup}}$ for HD 984 B, placing the companion firmly in the brown dwarf regime. The new fit also reveals a higher eccentricity for the companion ($e = 0.76 \pm 0.05$) compared to previous orbit fits. Given the broad age constraint for HD 984, this mass is consistent with predictions from evolutionary models. HD 984 B's dynamical mass places it among a small but growing list of giant planet and brown dwarf companions with direct mass measurements.
\end{abstract}

\keywords{Brown dwarfs (185), Direct imaging (387), Orbit determination (1175), Radial velocity (1332), Astrometry (80)}

\section{Introduction}
The masses of directly imaged brown dwarfs and giant planets are traditionally inferred from  low-temperature cooling models (e.g., \citealt{burrows_nongray_1997,saumon_evolution_2008, phillips_new_2020}). Unlike stars, substellar objects never reach sufficient core temperatures to sustain hydrogen burning. These objects instead lose energy and grow dimmer over their entire lifespans \citep{kumar_structure_1963, burrows_theory_2001}. Masses for these objects can then be inferred from their luminosity and age using evolutionary models. These inferred masses are generally uncertain and model-dependent (e.g., \citealt{bowler_imaging_2016}). For instance, the mass of the giant planet 51 Eridani b ranges from $1{-}12$ $\mathrm{M_{Jup}}$ depending on the model choice \citep{macintosh_discovery_2015, nielsen_gemini_2019}. Similarly, inferred masses of the wide-separation companion COCONUTS-2b can vary from $\sim 6{-}12$ $\mathrm{M_{Jup}}$ \citep{zhang_second_2021}.

Several assumptions are made in constructing substellar evolutionary models. A significant parameter, particularly for young objects, is the amount of entropy the planet or brown dwarf starts with immediately post-formation. Objects following hot-start models (e.g., \citealt{burrows_nongray_1997, baraffe_evolutionary_2003}) begin with arbitrarily large radii and high entropy. Eventually, as hot-start objects cool and contract on a Kelvin-Helmholz timescale, the impact of this initial condition fades. Hot-start grids are thought to better represent formation via gravitational instabilities in a disk or molecular cloud \citep{spiegel_spectral_2012, chabrier_giant_2014} and are likely to be more applicable to brown dwarf companions than giant planets (e.g, \citealt{kratter_runts_2010}). Cold-start models strive to more accurately emulate giant planet formation through core accretion by removing initial internal energy (e.g., \citealt{marley_luminosity_2007, fortney_synthetic_2008}). This reflects the dissipation of energy through accretion shocks and results in much lower luminosities (and thus higher inferred masses of imaged planets) at young ages. Warm-start models (e.g., \citealt{spiegel_spectral_2012, mordasini_luminosity_2013, marleau_constraining_2014}) strike a middle ground between these two classes of models. With intermediate initial entropies, the goal of these models is to portray a more realistic core accretion scenario, with less accretion luminosity lost to radiation than their cold-start counterparts. However, significant uncertainties remain about the initial entropy content of objects that form through these processes. The luminosities of young objects of a given mass and age differ substantially among these three classes of models, resulting in a degeneracy between the inferred mass of a planet or brown dwarf and the formation pathway. 

Additional considerations also affect the evolution of substellar objects in these models. The treatment of cloud formation and dissipation can influence giant planet and brown dwarf cooling by modifying their atmospheric opacity. Furthermore, the compositions of atmospheres can impact substellar evolution. This is a subtle effect, however, due to the weak dependence of luminosity on mean opacity \citep{burrows_science_1993}. Evolutionary models can also differ in the H/He equations of state for modeling high-density matter in brown dwarf and giant planet interiors \citep[e.g.,][]{saumon_equation_1995, chabrier_new_2019}. As brown dwarfs contract, they reach sufficient temperatures to induce deuterium burning (along with lithium burning for the highest-mass substellar objects; e.g., \citealt{gharib-nezhad_following_2021}). The energy generated during this process is not enough to balance radiative losses, but it results in a slight delay in cooling. The specific details of the onset and eventual cessation of deuterium and lithium burning at a given mass are dependent on metallicity, helium fraction, and initial entropy \citep{spiegel_deuterium-burning_2011}. This serves as an additional source of variation among brown dwarf cooling models.

\begin{figure*}
    \centering
    \includegraphics[width=0.7\textwidth]{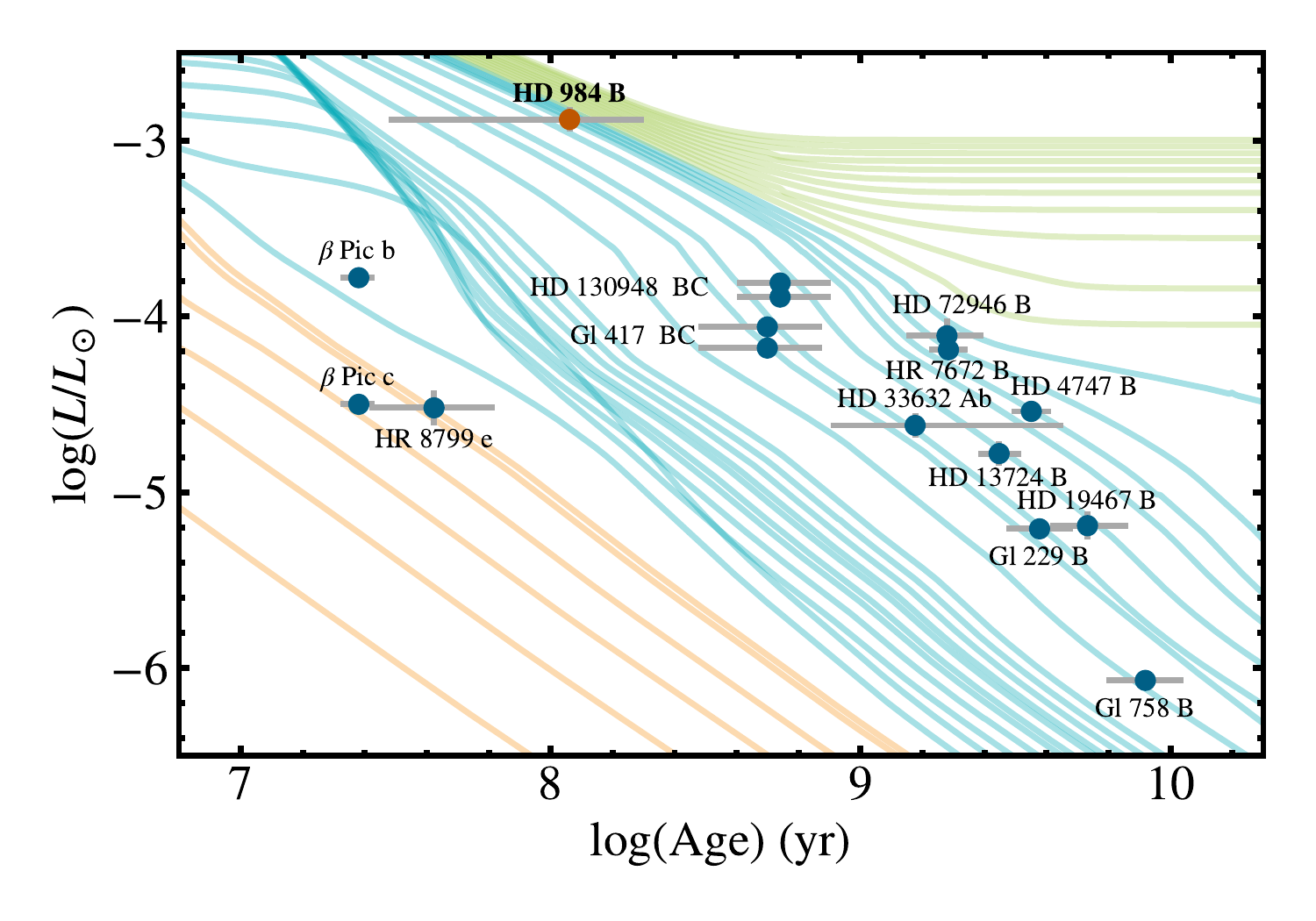}
    \vskip -0.1in
    \caption{Imaged substellar companions with dynamical masses and bolometric luminosities. Cooling tracks of stars, brown dwarfs, and giant planets from \citet{burrows_nongray_1997} are shown in green, blue, and orange, respectively. HD 47127 B \citep{bowler_mcdonald_2021-1} and HD 4113 C \citep{cheetham_direct_2018} are not included in this plot due to their lack of published bolometric luminosities. The eclipsing brown dwarf binaries 2MASS J05352184-0546085 \citep{stassun_discovery_2006} and 2MASS J15104786-2818174 Aab \citep{triaud_eclipsing_2020}, along with the transiting brown dwarf companion CWW 89Ab \citep{beatty_significant_2018}, have measured dynamical masses but have not been spatially resolved. Unlike most brown dwarfs with dynamical masses, HD 984 B is young and offers an opportunity to assess substellar evolutionary models in an untested region of luminosity-age space near the substellar boundary.
    \label{fig:dm_lum_age}}
\end{figure*}

\begin{deluxetable*}{lcccc}
\tablecaption{\label{tab:dm}Imaged Substellar Companions with Dynamical Masses}
\tablehead{\colhead{Companion} & \colhead{Dynamical Mass} & \colhead{Age} & \colhead{Luminosity} & \colhead{Refs.}\\
\colhead{ } & \colhead{($\mathrm{M_{Jup}}$)} & \colhead{(Myr)} & \colhead{($\log(L_{\mathrm{bol}} / \mathrm{L_\odot})$)} & \colhead{ }}
\startdata
$\beta$ Pic c &$8.2 \pm 1.0$ &$24.0 \pm 3.0$ &$-4.50 \pm 0.04$ &1, 2, 3, 4\\
$\beta$ Pic b &$9.3 \pm 2.6$ &$24.0 \pm 3.0$ &$-3.780 \pm 0.030$ &1, 2, 5\\
HR 8799 e &$9.6 \pm 1.9$ &$42 \pm 24$ &$-4.52 \pm 0.10$ &6\\
HD 13724 B &$36.2 \pm 1.6$ &$2800 \pm 500$ &$-4.78 \pm 0.07$ &7\\
Gl 758 B &$38.0 \pm 0.8$ &$8300 \pm 2700$ &$-6.070 \pm 0.030$ &7, 8\\
HD 33632 Ab &$50 \pm 6$ &$1700 \pm 400$ &$-4.62 \pm 0.07$ &7\\
$\mathsf{\varepsilon}$ Indi C &$53.12 \pm 0.32$\tablenotemark{a} & $\cdots$ &$-5.232 \pm 0.020$ &9, 10\\
HD 984 B &$61 \pm 4$ &30--200 &$-2.87 \pm 0.07$ &11, 12, 13\\
HD 19467 B &$65 \pm 6$ &$5400 \pm 1900$ &$-5.19 \pm 0.08$ &7\\
HD 4113 C &$66 \pm 5$ &$5000 \pm 1300$ & $\cdots$ &14\\
Gl 802 B &$66 \pm 5$ &${\gtrsim}1000$ &$-4.54$ &15\\
HD 4747 B &$66.2 \pm 2.5$ &$3600 \pm 600$ &$-4.54 \pm 0.06$ &16\\
$\mathsf{\varepsilon}$ Indi B &$68.0 \pm 0.9$\tablenotemark{a} & $\cdots$ &$-4.699 \pm 0.017$ &9, 10\\
Gl 229 B &$71.4 \pm 0.6$ &$3800 \pm 1100$ &$-5.208 \pm 0.007$ &7, 17, 18\\
HD 72946 B &$72.5 \pm 1.3$ &$1900 \pm 600$ &$-4.133 \pm 0.023$ &7\\
HR 7672 B &$72.7 \pm 0.8$ &$1920 \pm 300$ &$-4.19 \pm 0.04$ &16\\
Gl 417 C &$99.0 \pm 3.0$\tablenotemark{b} &300--750 &$-4.18 \pm 0.04$ &19, 20\\
Gl 417 B &$99.0 \pm 3.0$\tablenotemark{b} &300--750 &$-4.06 \pm 0.04$ &19, 20\\
HD 47127 B &$105 \pm 18$ &7000--10000 & $\cdots$ &21\\
HD 130948 C &$114.0 \pm 3.0$\tablenotemark{b} &400--800 &$-3.890 \pm 0.030$ &19, 20, 22\\
HD 130948 B &$114.0 \pm 3.0$\tablenotemark{b} &400--800 &$-3.810 \pm 0.030$ &19, 20, 22\\
\enddata
\tablenotetext{a}{Note that \citet{dieterich_dynamical_2018} measure masses of $75.0 \pm 0.8$ $\mathrm{M_{Jup}}$ and $70.1 \pm 0.7$ $\mathrm{M_{Jup}}$ for $\mathsf{\varepsilon}$ Indi B and C, respectively, which differ from the values from \citet{cardoso_observational_2012} listed here.}
\tablenotetext{b}{Both Gl 417 BC and HD 130948 BC are brown dwarf binaries in hierarchical triple systems with a wider separation primary. The masses listed for these companions are the total masses of each brown dwarf binary.}
\tablerefs{(1) \citet{brandt_precise_2021}; (2) \citet{bell_self-consistent_2015}; (3) \citet{nowak_direct_2020}; (4) Mathias Nowak, private communication; (5) \citet{morzinski_magellan_2015}; (6) \citet{brandt_first_2021}; (7) \citet{brandt_improved_2021}; (8) \citet{bowler_orbit_2018}; (9) \citet{cardoso_observational_2012}; (10) \citet{king_mathsfvarepsilon_2010}; (11) This work; (12) \citet{meshkat_discovery_2015}; (13) \citet{johnson-groh_integral_2017}; (14) \citet{cheetham_direct_2018}; (15) \citet{ireland_dynamical_2008}; (16) \citet{brandt_precise_2019}; (17) \citet{brandt_dynamical_2020}; (18) \citet{filippazzo_fundamental_2015}; (19) \citet{dupuy_new_2014}; (20) \citet{vigan_vltnaco_2017}; (21) \citet{bowler_mcdonald_2021-1}; (22) \citet{dupuy_dynamical_2009}}
\end{deluxetable*}

\begin{deluxetable}{lcc}
\tablecaption{\label{tab:prop}Properties of HD 984 A and B}
\tablehead{\colhead{Property} & \colhead{Value} & \colhead{Refs}}
\startdata
\multicolumn{3}{c}{Host Star} \\
\hline
$\alpha_{2000.0}$ & 00:14:10.25 & 1\\
$\delta_{2000.0}$ & $-07$:11:56.81 & 1\\
$\pi$ (mas) & $21.877 \pm 0.025$ & 1\\
Distance (pc) & $45.71 \pm 0.05$ & 1\\
SpT & F7 V & 2\\
Mass ($\mathrm{M_\odot}$) & $1.2 \pm 0.01$ & 3, 4, 5\\
Age (Myr) & $30{-}200$ & 3\\
$T_{\mathrm{eff}}$ (K) & 6315 & 6\\
$\mathrm{[Fe/H]}$ (dex) & ${+}0.27$ & 7\\
$v \sin i$ ($\mathrm{km/s}$) & $39.3 \pm 1.5$ & 8\\
$\log(R'_{HK})$ (dex) & $-4.34$ & 9\\
$\log(L_X/L_{\mathrm{bol}})$ (dex) & -4.22 & 3, 10\\
RUWE & 0.976 & 1\\
$B$ (mag) & $7.82 \pm 0.02$ & 11\\
$V$ (mag) & $7.32 \pm 0.01$ & 11\\
Gaia $G$ (mag) & $7.2077 \pm 0.0028$ & 1\\
$J$ (mag) & $6.402 \pm 0.023$ & 12\\
$H$ (mag) & $6.17 \pm 0.04$ & 12\\
$K_s$ (mag) & $6.073 \pm 0.021$ & 12\\
\hline
\multicolumn{3}{c}{Companion} \\
\hline
Mass ($\mathrm{M_{Jup}}$) & $61 \pm 4$ & 13\\
SpT & M$6.5\pm 1.5$ & 14\\
$T_{\mathrm{eff}}$ (K) & $2730^{+120}_{-180}$ & 14\\
$\log(L_{\mathrm{bol}} / L_\odot)$ (dex) & $-2.88 \pm 0.03$ & 13\\
$J$ (mag) & $13.28 \pm 0.06$ & 14\\
$H$ (mag) & $12.60 \pm 0.05$ & 14\\
$K_s$ (mag) & $12.20 \pm 0.04$ & 13\\
Semi-Major Axis (AU) & $28^{+7}_{-4}$ & 13\\
Eccentricity & $0.76 \pm 0.05$ & 13\\
Inclination (${}^{\circ}$) & $120.8^{+1.8}_{-1.6}$ & 13\\
Period (yr) & $140^{+50}_{-30}$ & 13\\
\enddata
\tablerefs{(1) \citet{gaia_collaboration_gaia_2020}; (2) \citet{white_high-dispersion_2007}; (3) \citet{meshkat_discovery_2015}; (4) \citet{mints_unified_2017}; (5) \citet{nielsen_gemini_2019}; (6) \citet{casagrande_new_2011}; (7) \citet{luck_abundances_2018}; (8) \citet{zuniga-fernandez_search_2021}; (9) \citet{boro_saikia_chromospheric_2018}; (10) \citet{voges_rosat_1999}; (11) \citet{hog_tycho-2_2000}; (12) \citet{skrutskie_two_2006}; (13) This work; (14) \citet{johnson-groh_integral_2017}}
\end{deluxetable}

Masses can be directly measured when the dynamical influence that a companion exerts on its host star can be observed. One way to achieve this is to couple observations of the \textit{relative} motion of the companion with \textit{absolute} astrometry of the host star. When accompanied by an age constraint and a luminosity measurement, dynamical masses offer a powerful means to empirically calibrate substellar evolutionary models. A comprehensive sample of these benchmark systems across a large dynamic range of age (1 Myr – 10 Gyr) and luminosity ($10^{-7}$ – $10^{-2}$ $L_\odot$) would facilitate detailed tests of substellar evolutionary models. These measurements are challenging to obtain, however, due to the long orbital periods of imaged companions. Only 15 dynamical masses have been measured for imaged substellar companions with well-constrained ages (shown in Figure \ref{fig:dm_lum_age} and Table \ref{tab:dm}). Each new system provides a valuable anchor to empirically map how giant planets and brown dwarfs cool over time.

HD 984 is a young ($30{-}200$ Myr) F7 star with a low-mass companion \citep{meshkat_discovery_2015}. Hot-start masses of the companion range from 34 to $\SI{94}{M_{Jup}}$, depending on the adopted system age. HD 984 B is located at a projected separation of ${\sim}200$ mas (${\sim}9.5$ AU), has a contrast of $\Delta H = 6.43\pm 0.05$ mag, and a spectral type of M$6.5\pm1.5$ \citep{meshkat_discovery_2015, johnson-groh_integral_2017}. Table \ref{tab:prop} summarizes the properties of the host star and companion. HD 984 B has been imaged with adaptive optics from $2012{-}2015$ with VLT/NaCo, VLT/SINFONI, and Gemini-S/GPI and shows orbital motion. Additionally, this star exhibits a significant proper motion difference between \emph{Hipparcos} and \emph{Gaia} in the \emph{Hipparcos}-\emph{Gaia} Catalog of Accelerations (HGCA; \citealt{brandt_hipparcos_2018, brandt_hipparcos-gaia_2021}).

Here, we measure the dynamical mass and 3D orbit of HD 984 B by jointly fitting the HGCA proper motions, relative astrometry, and radial velocities of the system. We obtained new Keck/NIRC2 imaging of the companion to extend the short orbital arc from previous studies as well as new precision radial velocities of the host star with the Habitable-Zone Planet Finder (HPF) spectrograph on the Hobby-Eberly Telescope to measure the radial acceleration of the host star. Section \ref{sec:observations} presents our imaging and radial velocity observations. In Section \ref{sec:orbitfit}, we fit the orbit of HD 984 B, which also constrains the companion's dynamical mass. In Section \ref{sec:discussion}, we compare our dynamical mass to predicted masses from substellar evolutionary models. Our results are summarized in Section \ref{sec:conclusion}.

\section{Observations \& Data Reduction \label{sec:observations}}
\subsection{Keck/NIRC2 Adaptive Optics Imaging}
We obtained high-contrast imaging of the HD 984 system with the NIRC2 camera at the W.M. Keck Observatory on UT 2019 July 7 and UT 2020 July 30 using natural guide star adaptive optics \citep{wizinowich_astronomical_2013}. Our observations consisted of short exposures in $K_s$-band with 500 coadded integrations and $t_{\mathrm{int}}=\SI{0.02}{s}$ per coadd. We obtained these images without a coronagraph due to the close separation of the companion from its host star (${\sim}200$ mas). To avoid saturating the detector with the unocculted primary, we read out a subarray of $192\times248$ pixels, which enables shorter integration times. The 2019 and 2020 datasets contained 176 and 120 exposures, respectively. This amounts to 29.3 min and 20 min of total integration time. All images were acquired in pupil-tracking mode to leverage the Angular Differential Imaging (ADI) method \citep{liu_substructure_2004, marois_angular_2006}. The 2019 and 2020 sequences spanned $30\fdg5$ and $21\fdg7$ of sky rotation, respectively.

The first step of our image reduction is subtracting dark frames and flat-fielding the images. Cosmic rays are identified and removed via the \texttt{L.A.Cosmic} algorithm \citep{vandokkum_cosmicray_2001}. We correct for geometric distortions by applying a piecewise affine transformation to each image using the solution from \citet{service_new_2016}. Frames are then registered by fitting a 2D Gaussian to HD 984, shifted so they are aligned to a common position. We perform this fit twice, which produces superior image alignment over a single iteration or alternatively cross-correlating the frames. Individual images are interpolated with sub-pixel precision using a 3rd-order spline during the alignment process.

We use the Vortex Image Processing (\texttt{VIP}) package \citep{gomez_gonzalez_vip:_2017} to conduct PSF subtraction. \texttt{VIP} offers several choices for PSF subtraction, including non-negative matrix factorization (NMF; \citealt{ren_non-negative_2018}), local low-rank plus sparse plus Gaussian-noise decomposition (LLSG; \citealt{gomez_gonzalez_low-rank_2016}), and a Principal Component Analysis (PCA) approach similar to the algorithms outlined in \citet{amara_span_2012} and \citet{soummer_detection_2012}. We adopt the PCA implementation because of its performance and speed. A lower-dimensional orthogonal basis of eigenimages is first constructed from the full image sequence. The eigenimages (principal components) are found by computing the singular value decomposition of the data cube. The reference PSF is then constructed for a given image by projecting the frame onto this basis. Reference PSFs are subsequently subtracted for each image and the reduced frames are derotated to a common angle and median combined, leaving an image with the signal of the companion added coherently and the residual background predominantly added incoherently.

\begin{figure}
    \centering
    \includegraphics[width=\linewidth]{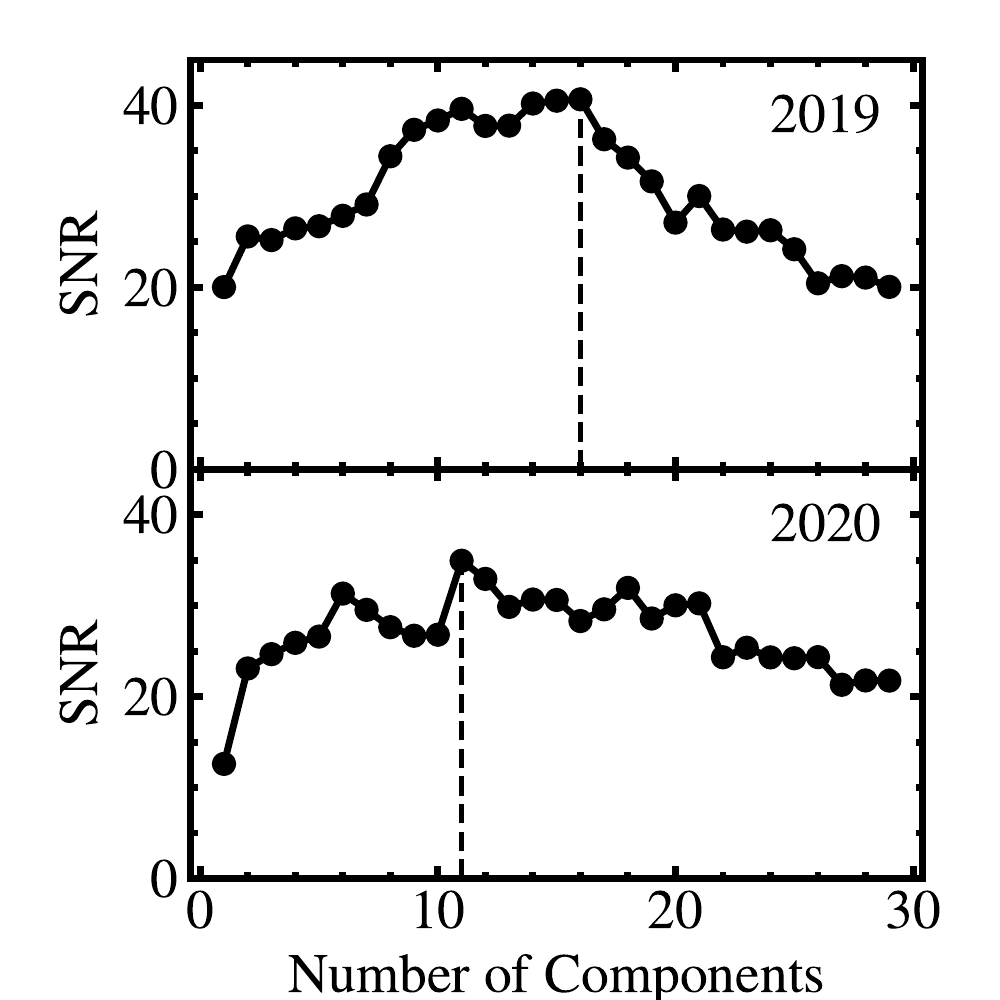}
    \caption{SNR of HD 984 B in PSF-subtracted NIRC2 images as a function of the number of PCA components. We find that a 16-component reduction  produces the highest SNR of 40.7 for the 2019 epoch. An 11-component reduction produces the highest SNR of 34.9 for the 2020 epoch. We thus adopt these numbers of components for our PSF subtraction.
    \label{fig:snr_comp_plot}}
\end{figure}

\begin{figure*}
    \centering
    \includegraphics[width=\textwidth]{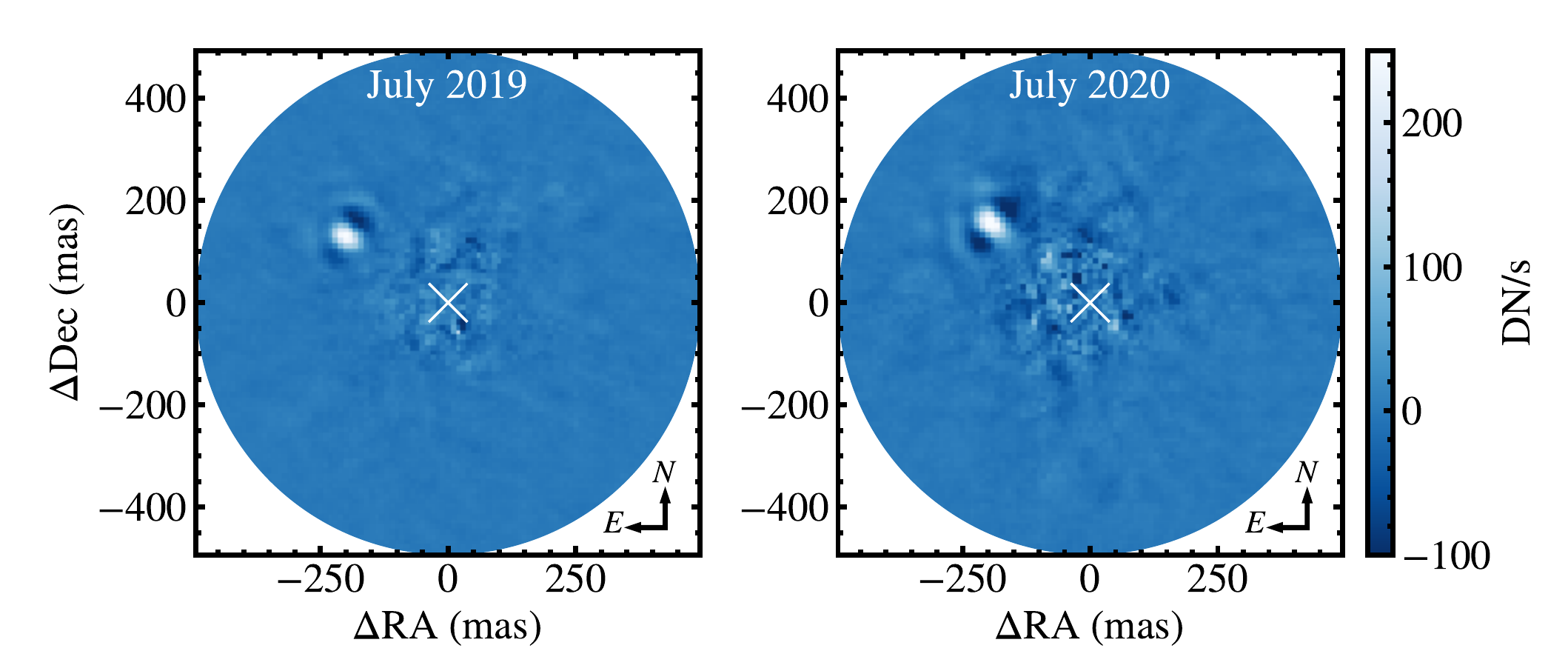}
    \caption{PSF-subtracted images of HD 984 B taken with NIRC2 in July 2019 (left) and July 2020 (right). The color bar corresponds to the flux in each reduced image in units of $\mathrm{DN/s}$. The images are oriented so north is up and east is to the left.
    \label{fig:keck_reduced}}
\end{figure*}

One advantage of a PCA-based approach is that tuning of the subtraction is largely confined to a single parameter: the number of principal components. Using too few components would undersubtract the host star PSF, while using too many components would subtract more of the companion's signal. To optimize the number of components, we run the PSF subtraction for each epoch with a total number of components ranging from 1 to 30. We then calculate the signal-to-noise ratio (SNR) for each subtracted image through aperture photometry within a 0.5 FWHM-radius circular aperture. The image coordinates of the companion are determined by fitting a 2D Gaussian to a 14-pixel wide subset of the image around the companion. This is done for each number of principal components. We estimate the noise-level by measuring the flux in non-overlapping apertures at the same angular radius from the star over a range of azimuthal angles excluding that of the companion. A maximum of 24 and 22 non-overlapping noise estimation apertures with radius 0.5 FWHM are used for the 2019 and 2020 epochs, respectively, based on the angular separation of HD 984 B in the images and excluding apertures within 5 aperture radii of the companion. The ratio of the companion flux and the standard deviation of the flux within the noise estimation apertures then yields the SNR. Figure \ref{fig:snr_comp_plot} shows the measured SNR for different numbers of components for each epoch. The highest companion SNRs are produced by PSF subtractions of 16 components and 11 components for the 2019 and 2020 epochs, respectively.\footnote{We evaluate the effect of the number of PCA components on the ensuing analysis by measuring astrometry and running the orbit fits for several alternative choices. This analysis is carried out with the second-highest SNR for each epoch (15 and 12 components for 2019 and 2020, respectively) and the third-highest SNR (14 and 18 components). The resulting astrometry is in good agreement with our adopted results, and the orbit fits from these two runs produce nearly identical values and the adopted uncertainties for the companion mass, semi-major axis, eccentricity, and inclination as the orbit fit with astrometry reduced using the highest SNR components.} The subtracted images for each observation are shown in Figure \ref{fig:keck_reduced}.

To avoid introducing bias in our astrometry from the PSF subtraction algorithm, we follow the approach of \citet{lagrange_giant_2010} and \citet{marois_images_2010} of injecting a negative-amplitude PSF template in the original, preprocessed frames at the approximate location of the companion. At the true position of the companion, and with the true flux, the signal of the companion in the final PSF-subtracted image will be removed. For the PSF template, we use the median combination of the images of HD 984 in our ADI sequence. The optimal companion separation ($\rho$), position angle ($\theta$), and flux ratio ($f_2 / f_1$) are measured by minimizing the signal within a circular aperture of 3 FWHM of the PSF template at the location of the companion in the processed image. The companion parameters are first coarsely optimized in the \texttt{VIP} package using the \texttt{AMOEBA} downhill simplex algorithm \citep{nelder_simplex_1965}. With this as a starting point, the parameter space is then finely explored using the \texttt{emcee} affine-invariant Markov chain Monte Carlo (MCMC) ensemble sampler \citep{foreman-mackey_emcee_2013}. We use 100 walkers to sample our three companion parameters ($\rho$, $\theta$, $f_2/f_1$) over ${\approx}3\times10^4$ steps. The number of steps is determined by the Gelman-Rubin test for convergence \citep{gelman_inference_1992}. We adopt the default threshold in \texttt{VIP} of $\hat{\mathcal{R}} < 1.01$, where $\hat{\mathcal{R}}$ is the Gelman-Rubin statistic which compares the variance between MCMC chains and the variance within individual chains. The flux ratio is then converted to an apparent magnitude of the companion using photometry of HD 984 from 2MASS ($K_s = 6.073 \pm 0.021$ mag; \citealt{skrutskie_two_2006}).

Figure \ref{fig:triangle_astrometry} shows joint parameter posterior distributions from the MCMC runs for our two epochs. Our new astrometry is listed alongside previous measurements in Table \ref{tab:rel_astrometry}. We measure contrasts of $\Delta K_s = 6.10 \pm 0.05$ mag for the 2019 epoch and $\Delta K_s = 6.15 \pm 0.05$ mag for the 2020 epoch. These correspond to apparent magnitudes of HD 984 B of $K_s = 12.17 \pm 0.05$ mag and $K_s = 12.22 \pm 0.05$ mag for each epoch, respectively. These compare well with the previous magnitude measurement from \citet{meshkat_discovery_2015} of $K_s = 12.19 \pm 0.04$ mag.

There are four dominant sources of uncertainty for our NIRC2 astrometry: errors in the measurement of the position of the companion, the distortion correction, the plate scale, and north alignment of the detector. Our measurement error is derived from the marginalized MCMC posteriors of the separation $\rho$ and position angle $\theta$. We average the upper and lower error bars of the central $\pm1\sigma$ (68.3\%) confidence interval of each posterior to produce the uncertainties $\sigma_{\mathrm{meas}, \rho}$ and $\sigma_{\mathrm{meas}, \theta}$. The raw images are corrected for optical distortion effects using the solution from \citet{service_new_2016}. They found that the post-correction residual uncertainties averaged across the detector are about $\sigma_{\mathrm{d}} = \SI{1}{mas}$. To convert the astrometry from units of pixels to angular units, we adopt the NIRC2 plate scale from \citet{service_new_2016}. Their plate scale $s$ and associated uncertainty $\sigma_{s}$ are $\SI{9.971(4)}{mas / pix}$. Finally, for the position angle, the \citet{service_new_2016} distortion correction provides the angle to align the images with celestial north: $\theta_{\mathrm{north}} = 0\fdg262 \pm 0\fdg002$.

The separation in angular units is calculated from the measured separation in units of pixels, $\rho_{\mathrm{meas}}$, via $\rho = s \rho_{\mathrm{meas}}$. The uncertainty on this separation is
\begin{equation}
    \sigma_{\rho} = s \rho_{\mathrm{meas}} \bigg(\Big(\frac{\sigma_s}{s}\Big)^2 + \frac{\sigma_{\mathrm{meas},\rho}^2 + 2 \sigma_d^2}{\rho_{\mathrm{meas}}^2}\bigg)^{1/2}.
\end{equation}
The distortion correction uncertainty is multiplied by $\sqrt{2}$ to account for measuring both the position of the host star (during image registration) and the companion (with the negative injection procedure).

The position angle of the companion is found from the uncorrected position angle $\theta_{\mathrm{meas}}$ through
\begin{equation}
    \theta = \theta_{\mathrm{meas}} - \theta_{\mathrm{VA}} -  \theta_{\mathrm{north}}.
\end{equation}
In vertical angle (pupil tracking) mode with NIRC2, $\theta_{\mathrm{VA}}$ is computed from header values in the following manner:
\begin{equation}
    \theta_{\mathrm{VA}} = \texttt{PARANG} + \texttt{ROTPOSN} - \texttt{INSTANGL},
\end{equation}
where \texttt{PARANG} is the parallactic angle, \texttt{ROTPOSN} is the rotator position, and \texttt{INSTANGL} is the NIRC2 position angle zeropoint of $0\fdg7$. The uncertainty on the position angle is the error from each term added in quadrature:
\begin{equation}
    \sigma_\theta = \sqrt{\sigma_{\mathrm{meas}, \theta}^2 + \sigma_{\mathrm{north}}^2 + \sigma_{d, \theta}^2}.
\end{equation}
The angular uncertainty from the distortion solution is $\sigma_{d, \theta} \approx \sigma_d / \rho$, which is derived by applying the small angle approximation.

\begin{figure}
    \centering
    \includegraphics[width=\linewidth]{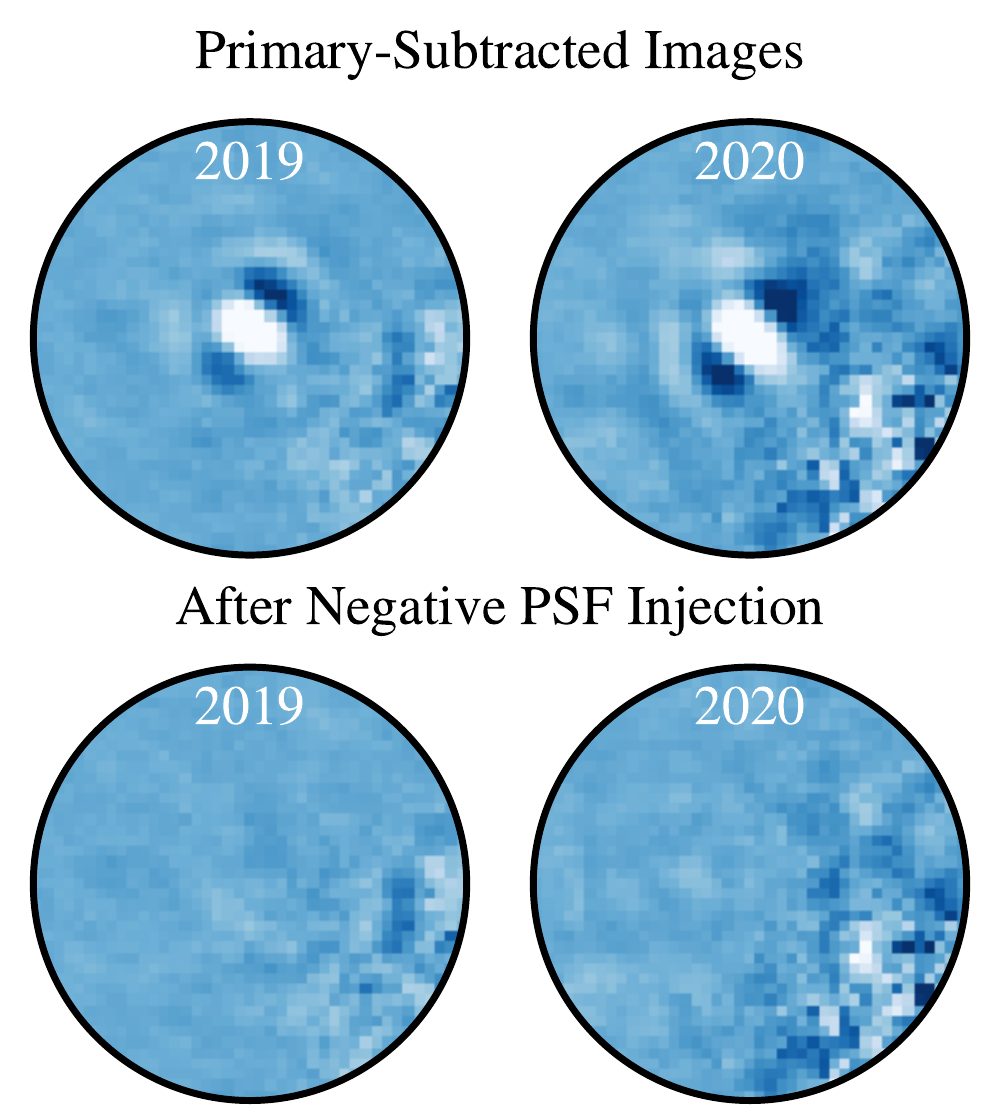}
    \caption{Results of our negative PSF injection. The top panels are the PSF-subtracted images of HD 984 B from our July 2019 and July 2020 NIRC2 datasets. The bottom panels show the results after injecting negative PSF templates with the best-fit separations, position angles, and flux ratios. The elevated noise in the lower-right portion of the images is due to poorer PSF-subtraction closer to the host star.
    \label{fig:neg_injection}}
\end{figure}

\begin{figure*}
    \centering
    \includegraphics[width=\textwidth]{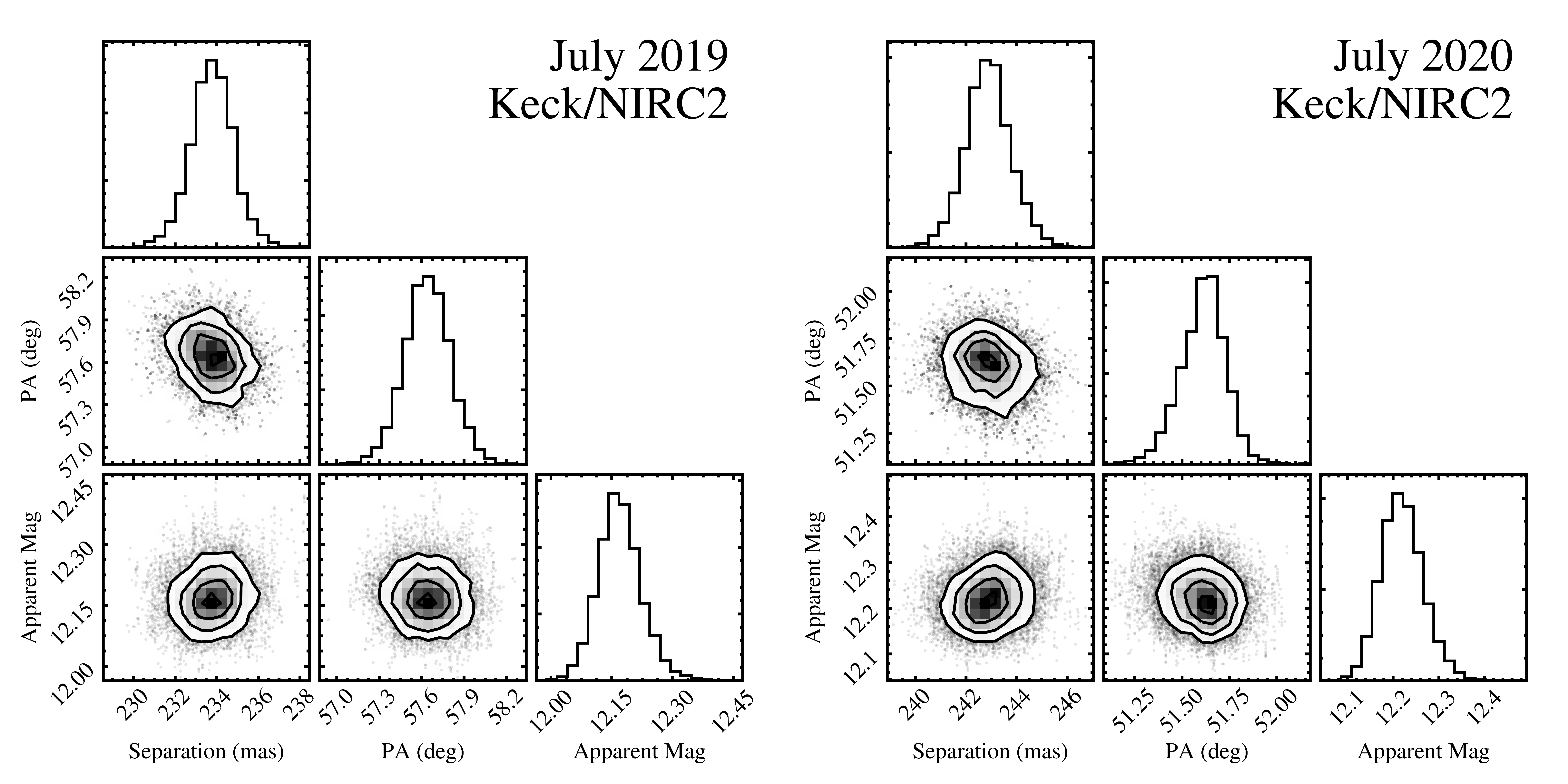}
    \caption{Posterior distributions of HD 984 B astrometry from our July 2019 (left) and July 2020 (right) NIRC2 datasets. The diagonal panels show the marginalized distributions for each parameter. The off-diagonal panels display two-dimensional joint posterior distributions with 1, 2, 3, and $4\sigma$ contours. Astrometry is measured via a negative companion injection approach that uses MCMC to explore the parameter space. All corner plots in this study are produced using \texttt{corner} \citep{foreman-mackey_cornerpy_2016}. 
    \label{fig:triangle_astrometry}}
\end{figure*}

\begin{deluxetable*}{lcccccc}
\tablecaption{\label{tab:rel_astrometry}Relative Astrometry of HD 984 B}
\tablehead{\colhead{Filter} & \colhead{Date} & \colhead{Epoch} & \colhead{Separation} & \colhead{PA} & \colhead{Instrument} & \colhead{Reference}\\ \colhead{ } & \colhead{(UT)} & \colhead{(UT)} & \colhead{(mas)} & \colhead{(\si{\degree})} & \colhead{ } & \colhead{ }}
\startdata
$L'$ & 2012 Jul 18 & 2012.545 & $190 \pm 20$ & $108.8 \pm 3.0$ & NaCo & \citet{meshkat_discovery_2015} \\
$L'$ & 2012 Jul 20 & 2012.550 & $208 \pm 23$ & $108.9 \pm 3.1$ & NaCo & \citet{meshkat_discovery_2015} \\
$H+K$ & 2014 Sep 10 & 2014.691 & $201.6 \pm 0.4$ & $92.2 \pm 0.5$ & SINFONI & \citet{meshkat_discovery_2015} \\
$H$ & 2015 Aug 30 & 2015.661 & $216.3 \pm 1.0$ & $83.3 \pm 0.3$ & GPI & \citet{johnson-groh_integral_2017} \\
$J$ & 2015 Aug 30 & 2015.661 & $217.9 \pm 0.7$ & $83.6 \pm 0.2$ & GPI & \citet{johnson-groh_integral_2017} \\
$K_s$ & 2019 Jul 07 & 2019.514 & $233.8 \pm 1.8$ & $57.64 \pm 0.29$ & NIRC2 & This Work \\
$K_s$ & 2020 Jul 30 & 2020.578 & $242.9 \pm 1.7$ & $51.61 \pm 0.26$ & NIRC2 & This Work
\enddata
\end{deluxetable*}

\subsection{Radial Velocities}
We obtained precision radial velocities of HD 984 with the Habitable-Zone Planet Finder (HPF), a high-resolution ($R \approx 55,000$) near-IR spectrograph on the 9.2-m Hobby-Eberly Telescope (HET) at McDonald Observatory \citep{mahadevan_habitable-zone_2012, mahadevan_habitable-zone_2014}. HPF possesses a laser-frequency comb (LFC) calibrator and active temperature stabilization at the milli-Kelvin level \citep{stefansson_versatile_2016} that together enable ${<}\SI{2}{m/s}$ precision on RV standards (e.g., \citealt{metcalf_stellar_2019, tran_epoch_2021}). The goal of these observations is to measure or constrain the radial component of the acceleration induced by HD 984 B on its host star. Measuring precise RVs of HD 984 is challenging due to its early spectral type (F7), fast rotation ($v\sin{i}=\SI{39.3(15)}{km/s}$; \citealt{zuniga-fernandez_search_2021}), and young age. Near-IR spectrographs like HPF have been shown to reduce activity-induced jitter by a factor of ${\approx} 2-3$ compared to optical spectrographs \citep{crockett_search_2012, gagne_high-precision_2016, tran_epoch_2021} by sampling wavelengths where starspot-to-photosphere contrasts are reduced.

We obtained a total of 13 RV epochs of HD 984 between UT 2019 September 11 and UT 2020 November 6 (Table \ref{tab:rv}). Observations were taken in queue mode using the HET's flexible scheduling system \citep{shetrone_ten_2007}. Each RV epoch of HD 984 consists of three 300-second exposures. For each exposure, 1D spectra are optimally extracted through the standard HPF data-extraction pipeline outlined in \citet{ninan_habitable-zone_2018} and \citet{metcalf_stellar_2019}. Radial velocities are measured using a least-squares matching technique based on the public \texttt{SERVAL} \citep{zechmeister_spectrum_2018} package as detailed in \citet{tran_epoch_2021}. All extracted 1D spectra are first corrected for Earth's barycentric motion using the \texttt{barycorrpy} \citep{kanodia_python_2018} software package, and high-frequency pixel-to-pixel variations are removed using a deblazed median master flat spectrum. 

All spectra are scaled to the highest SNR spectrum using multiplicative 3rd-order polynomials. This adjusts the spectra to the same flux level on a pixel-to-pixel basis and accounts for variations between observations. The master science template is then created by applying a uniform cubic basic regression to the normalized spectra. The best-fit spline coefficients are calculated by minimizing the residuals between the template and science spectra using a weighted $\chi^2$ value. The template undergoes $5\sigma$-clipping to remove points with high residuals; values near the template edge are ignored to account for any edge effects. Telluric absorption and sky emission lines are down-weighted during this process.

The radial velocity shifts and multiplicative polynomials are simultaneously fit by iteratively stepping through a grid of velocities and polynomial coefficients and comparing the template and each science spectrum. The master science template is Doppler shifted in steps of $\Delta v = 50$ m s$^{-1}$ from $-5$ km s$^{-1}$ to 5 km s$^{-1}$ and the $\chi^{2}$ residuals are calculated between this shifted template and each science spectrum. During this step, spectral regions in the science spectrum affected by telluric absorption and sky emission lines are masked. The final RV is taken to be the global minimum of the $\chi^{2}-$velocity curve. This process is carried out for 8 spectral orders in the HPF wavelength range least contaminated by telluric absorption. The final RV for a single exposure is the weighted mean of these spectral order RVs, and the standard error is adopted for the final uncertainty. Individual RVs for the contiguous three exposures are then binned together using a weighted mean to produce our RV measurement for a given epoch.

\begin{figure}
    \centering
    \includegraphics[width=\linewidth]{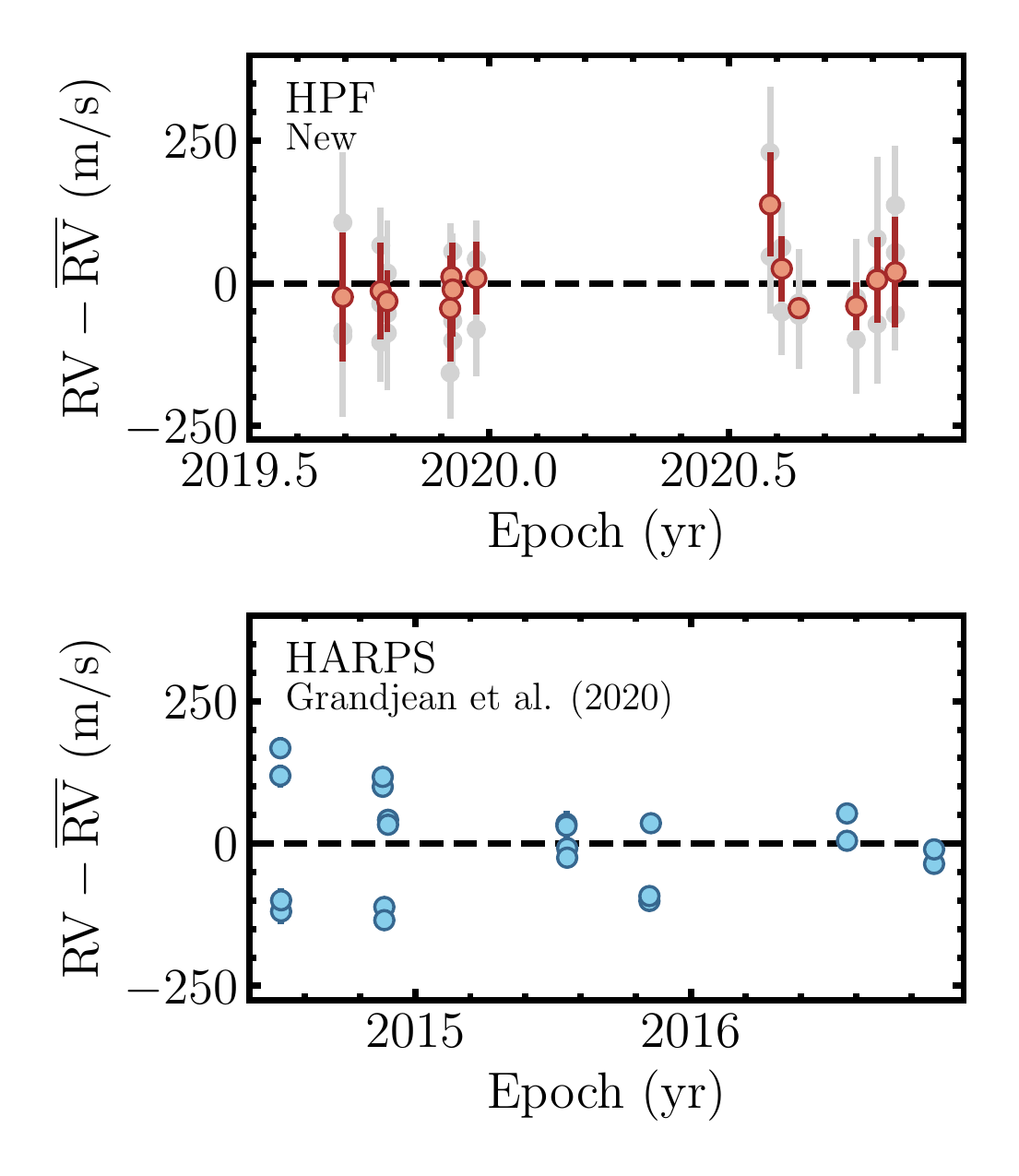}
    \caption{Relative radial velocities of HD 984 from HPF (top panel; this work) and HARPS (bottom panel; \citealt{grandjean_harps_2020}). HPF RVs from individual exposures are shown in gray, while the final, binned measurements for each visit are shown in red. No clear trends are observed, which may in part be caused by large scatter owing to HD 984's young age and early spectral type. Nevertheless, their inclusion in the orbit fit places constraints on the radial acceleration induced by HD 984 B.
    \label{fig:rv_obs_plot}}
\end{figure}

In addition to our HPF radial velocities, we incorporate 21 HARPS RVs in our analysis, which were obtained between UT 2014 July 6 and UT 2016 November 19 (\citealt{grandjean_harps_2020}; A. Grandjean, private communication). HARPS is a high-resolution ($R \approx 115,000$) optical spectrograph \citep{mayor_setting_2003} mounted at the 3.6-m ESO telescope at La Silla Observatory. Figure \ref{fig:rv_obs_plot} shows our new HPF RVs and the published HARPS RVs. The dispersion of the HPF RVs (48.0 m/s) is lower than that of the HARPS measurements (86.7 m/s) by a factor of 1.8. Nevertheless, the scatter is large in both datasets owing to both the intrinsic activity of HD 984 and the difficulty of measuring precise RVs of F stars, which have fewer spectral features and high rotational velocities compared to later-type stars.

Following \citet{tran_epoch_2021}, one way to assess the dominant sources of jitter for the two datasets is by considering the metric $\overline{\sigma}/\mathrm{RMS}$, where $\overline{\sigma}$ is the mean measurement uncertainty and $\mathrm{RMS}$ is the root mean square of the RV dataset. If this quantity is small (${<}1.0$), then the jitter is dominated by sources not incorporated into measurement uncertainties, indicating activity as the principal source. If this quantity is larger than 1.0, the RV scatter is captured by the instrumental uncertainties, indicating that spectral features (such as high $v \sin i$ and early spectral type) are the dominant origin of the RV scatter. For the HPF RVs, $\overline{\sigma}/\mathrm{RMS} = 1.5$, while for HARPS, $\overline{\sigma}/\mathrm{RMS} = 0.2$. This suggests that HD 984's high $v\sin i$ or broad spectral features are the principal source of jitter for the HPF data, while its spot-induced activity is the dominant source for the HARPS data. This is expected for RVs taken in the near-IR versus the optical.

The presence of an additional close-in stellar or substellar companion could also produce large RV variations on short timescales. We analyzed the HPF and HARPS datasets with a Lomb-Scargle periodogram to identify any periodicity present in the RVs that could correspond to a close-in companion. The periodograms for each instrument do not contain any signals above a false-alarm probability threshold of $0.1\%$.

Another way to assess the presence of close-in, massive companions is with the Renormalised Unit Weight Error (RUWE; \citealt{lindegren_re-normalising_2018}) statistic in \emph{Gaia} EDR3, which characterizes the goodness-of-fit for a single-star astrometric solution. RUWE values above ${\sim}1.4$ (and to an extent, above 1; see \citealt{stassun_parallax_2021}) correlate strongly with the presence of an unresolved binary. HD 984 has an RUWE value of 0.976, indicating that it is well-fit by a single-star solution. 

\begin{deluxetable}{lcr} 
\tablecaption{\label{tab:rv}HPF Radial Velocities}
\tablehead{\colhead{Date} & \colhead{RV} & \colhead{$\sigma_{\mathrm{RV}}$}\\
\colhead{($\mathrm{BJD_{TDB}}$)} & \colhead{($\mathrm{m/s}$)} & \colhead{($\mathrm{m/s}$)}}
\startdata
$2458737.842$ & $-30$ & $110$\\
$2458766.750$ & $-10$ & $90$\\
$2458771.741$ & $-30$ & $50$\\
$2458819.619$ & $-40$ & $90$\\
$2458820.607$ & $10$ & $3$\\
$2458821.596$ & $-10$ & $80$\\
$2458839.550$ & $10$ & $60$\\
$2459063.943$ & $140$ & $90$\\
$2459072.917$ & $20$ & $60$\\
$2459085.878$ & $-45$ & $11$\\
$2459129.754$ & $-40$ & $40$\\
$2459145.732$ & $10$ & $80$\\
$2459159.664$ & $20$ & $100$
\enddata
\end{deluxetable}

\subsection{Hipparcos-Gaia Acceleration}
HD 984 exhibits a significant proper motion difference (astrometric acceleration) between \emph{Hipparcos} and \emph{Gaia} which enables the measurement of its dynamical mass. Ensuring that these two astrometric datasets are appropriately cross calibrated is critical for comparing their proper motions. This was recently carried out in the construction of the \emph{Hipparcos}-\emph{Gaia} Catalog of Accelerations (HGCA; \citealt{brandt_hipparcos_2018}). The HGCA provides calibrated proper motions based on corrections of the \emph{Hipparcos} astrometry to the \emph{Gaia} frame through local rotations. It also incorporates a $40\%/60\%$ weighting of two \emph{Hipparcos} reductions \citep{esa_hipparcos_1997, van_leeuwen_validation_2007}, inflates uncertainties by fitting an additive error inflation term for \emph{Hipparcos} and a multiplicative error inflation factor for \emph{Gaia}, and cross-validates these steps. The result is a robust collection of proper motions with the sensitivity and time baseline (${\sim}25$ years) to enable the measurement of substellar dynamical masses. This has been recently carried out for imaged brown dwarf, giant planet, and white dwarf companions (e.g., \citealt{dupuy_model-independent_2019, brandt_precise_2019, brandt_dynamical_2020, brandt_precise_2021, currie_scexaocharis_2020, bowler_mcdonald_2021-1, bowler_mcdonald_2021}). 

Recently, \citet{brandt_hipparcos-gaia_2021} updated the HGCA to incorporate proper motions from \emph{Gaia} EDR3 \citep{gaia_collaboration_gaia_2020}. This greatly increases sensitivity to the low-amplitude proper motion changes that substellar companions induce on their host stars. For instance, the significance of the proper motion change for HD 984 increases from $20\sigma$ in the \emph{Gaia} DR2 version to $62\sigma$ in the \emph{Gaia} EDR3-equipped HGCA. We use these new cross-calibrated proper motions in the following analysis.

\begin{figure}
    \centering
    \includegraphics[width=\linewidth]{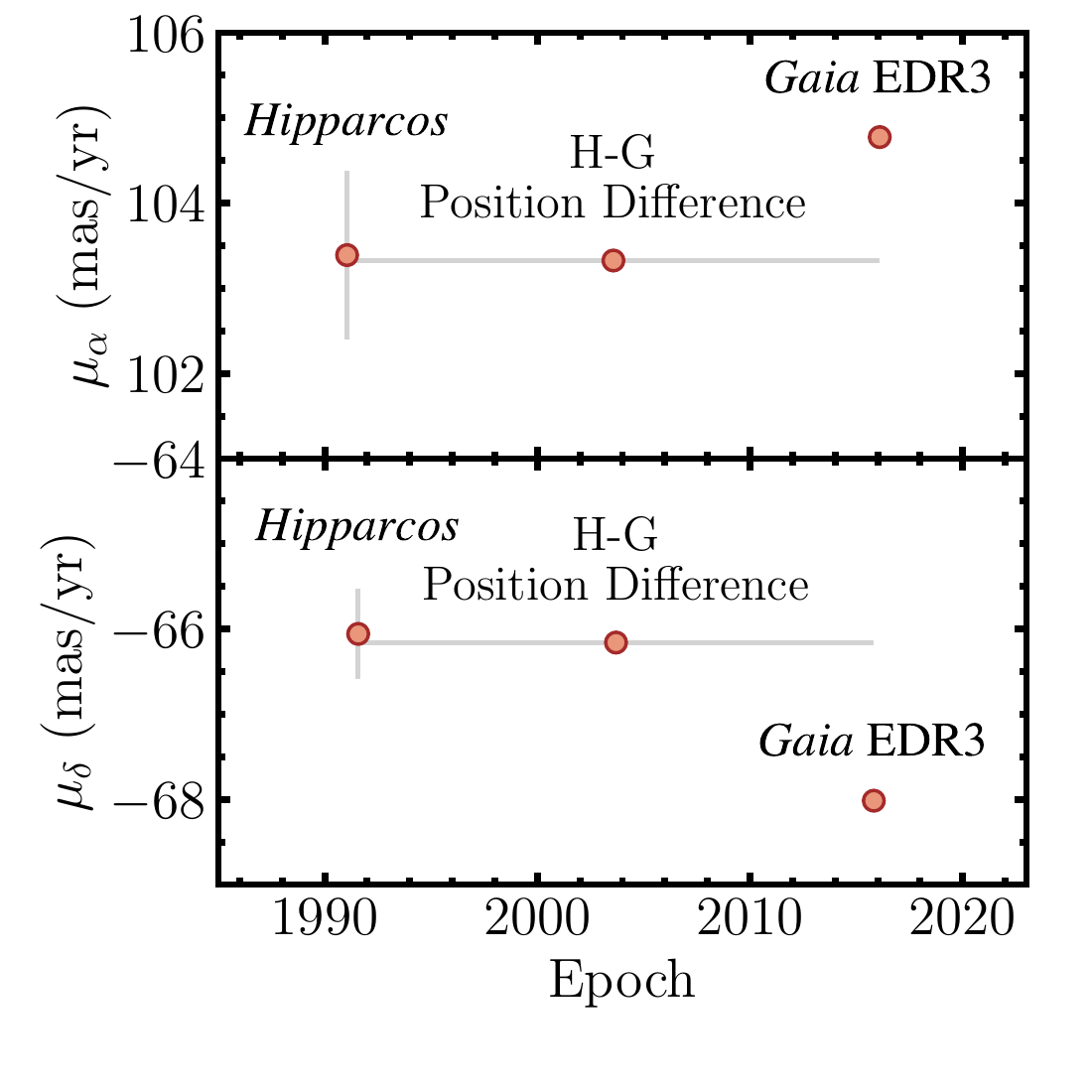}
    \vskip -0.1in
    \caption{HGCA proper motion measurements of HD 984 in R.A. (top) and Dec. (bottom). Three proper motions are available from the HGCA: the individual \emph{Hipparcos} and \emph{Gaia} proper motions and an average proper motion from the difference in sky position between the two missions. For HD 984, the position difference proper motion is the most precise, followed by the \emph{Gaia} epoch. The slope of these points indicates the time-averaged tangential (sky-plane) acceleration induced by the star's companion.
    \label{fig:hgca_pm}}
\end{figure}

Combining \emph{Hipparcos} and \emph{Gaia} astrometry produces three nearly independent proper motions: the individual proper motions at the \emph{Hipparcos} and \emph{Gaia} epochs and an average proper motion from the difference in sky position between the two missions. Figure \ref{fig:hgca_pm} shows these proper motions for HD 984. The measurement derived from the position difference is the most precise, followed by the \emph{Gaia} EDR3 proper motion. Table \ref{tab:hgca} lists the absolute HGCA astrometry for HD 984. This star exhibits a tangential acceleration between the \emph{Gaia} and long-term \emph{Hipparcos}-\emph{Gaia} proper motions of $41.6\pm 0.8$ m/s/yr.

\begin{deluxetable}{lr} 
\tablecaption{\label{tab:hgca}HGCA\tablenotemark{a} Kinematics for HD 984}
\tablehead{\colhead{Parameter} & \colhead{Value}}
\startdata
$\mu_{\alpha,\mathrm{Hip}}$\tablenotemark{b} (mas/yr) & $103.4 \pm 1.0$ \\
$\mu_{\alpha,\mathrm{Hip}}$ Epoch (yr) & 1991.05 \\
$\mu_{\delta,\mathrm{Hip}}$ (mas/yr) & $-66.1 \pm 0.5$ \\
$\mu_{\delta,\mathrm{Hip}}$ Epoch (yr) & 1991.57 \\
$\mu_{\mathrm{Hip}}$ Correlation Coefficient & 0.1119 \\
$\mu_{\alpha,\mathrm{Gaia}}$\tablenotemark{b} (mas/yr) & $104.77 \pm 0.05$ \\
$\mu_{\alpha,\mathrm{Gaia}}$ Epoch (yr) & 2016.1 \\
$\mu_{\delta,\mathrm{Gaia}}$ (mas/yr) & $-68.016 \pm 0.030$ \\
$\mu_{\delta,\mathrm{Gaia}}$ Epoch (yr) & 2015.82 \\
$\mu_{\mathrm{Gaia}}$ Correlation Coefficient & 0.0384 \\
$\mu_{\alpha,\mathrm{HG}}$\tablenotemark{b} (mas/yr) & $103.326 \pm 0.027$ \\
$\mu_{\delta,\mathrm{HG}}$ (mas/yr) & $-66.159 \pm 0.018$ \\
$\mu_{\mathrm{HG}}$ Correlation Coefficient & 0.269 \\
$\chi^2$ & 3810.4 \\
\enddata
\tablenotetext{a}{These measurements are from an updated version of the HGCA \citep{brandt_hipparcos-gaia_2021} that incorporates \emph{Gaia} EDR3.}
\tablenotetext{b}{Proper motions in R.A. include a factor of $\cos \delta$.}
\end{deluxetable}

\section{Orbit Fitting \label{sec:orbitfit}}
Our orbit fit incorporates relative astrometry from NaCo, SINFONI, GPI, and NIRC2; radial velocities from HARPS and HPF; and the HGCA proper motions. Astrometry from different instruments and post-processing routines can introduce systematic errors which can bias orbit fits. For example, \citet{bowler_population-level_2020} found that the P.A. measurements of the brown dwarf companion DH Tau B had additional scatter not captured by previously reported uncertainties. They attributed this to instrument-to-instrument errors in the distortion correction, north alignment, and/or plate scale. Based on an initial orbit fit of just the relative astrometry with \texttt{orbitize!} \citep{blunt_orbitize_2020}, we found that the astrometric uncertainties of HD 984 B are likely underestimated: the reduced $\chi^2$ value of the best-fit orbit is 5.8. This led us to devise an approach to inflate the uncertainties on the relative astrometry for the subsequent joint orbit fit using all available data.
\clearpage
\subsection{Astrometric Error Inflation}
Our approach to inflate the astrometric errors is to adopt an astrometric ``jitter" term $\sigma_{\mathrm{jit}}$. This term reflects additional positional uncertainty likely resulting from systematics among different instruments. This is added in quadrature to the reported separation uncertainties for our orbit fit:
\begin{align}
    \sigma_{\rho, \mathrm{fit}} = \sqrt{\sigma_\rho^2 + \sigma_{\mathrm{jit}}^2}.
    \label{eq:ei_sep}
\end{align}
The error on the position angle is inflated using this positional jitter term as follows:
\begin{align}
    \sigma_{\theta, \mathrm{fit}} = \sqrt{\sigma_\theta^2 + \bigg(\frac{\sigma_{\mathrm{jit}}}{\rho}\bigg)^2}.
    \label{eq:ei_pa}
\end{align}
We use \texttt{orbitize!} to find the value of $\sigma_{\mathrm{jit}}$ that produces appropriate error inflation that we would expect for normal standard deviates. \texttt{orbitize!} implements both ``Orbits for the Impatient" (\texttt{OFTI}), a Bayesian rejection-sampling approach to orbit fitting detailed in \citet{blunt_orbits_2017}, and an MCMC approach using the parallel-tempering Markov chain Monte Carlo (PT-MCMC) in \texttt{emcee} \citep{foreman-mackey_emcee_2013}. For the purposes of error inflation, we use the MCMC approach, as it is generally faster for well-constrained orbits. PT-MCMC involves running chains with several different temperatures, where higher-temperature chains more effectively escape local minima and lower-temperature chains more accurately sample peaks in the posterior distribution. Periodically, chains of different temperatures swap positions, enabling the colder ones to be sensitive to additional peaks present in the posterior accessed by higher-temperature chains. This approach is better suited to sampling the complex multi-model posteriors that can arise when fitting orbits. We used a total of 20 temperatures for each orbit fit, adopting the coldest set of chains for the posterior distribution. We experimented with different numbers of steps and found that a total of $10^5$ steps over $100$ walkers sampled the parameter space well. We discard the first $50\%$ of each chain as burn-in.

For each jitter value, we compute the reduced Chi-squared value $\chi^2_\nu$ through
\begin{equation}
    \chi^2_\nu = \frac{1}{\nu}\bigg( \sum_i \frac{(\rho_i - M_{\rho,i})^2}{\sigma_{\rho,\mathrm{fit},i}^2} + \sum_i \frac{(\theta_i - M_{\theta, i})^2}{\sigma_{\theta, \mathrm{fit}, i}^2} \bigg ),
\end{equation}
where $\nu$ is the number of degrees of freedom; $\rho_i$ and $\theta_i$ are the separation and position angle measurement, respectively; and $M_{\rho, i}$ and $M_{\theta, i}$ are the corresponding model values for the separation and position angle. The maximum likelihood orbit is used for the model to compute $\chi^2_\nu$. The number of degrees of freedom is the difference between the number of datapoints (16; seven measurements of the separation and position angle, plus the parallax and system mass) and the eight fitted parameters.

\begin{figure}
    \centering
    \includegraphics[width=\linewidth]{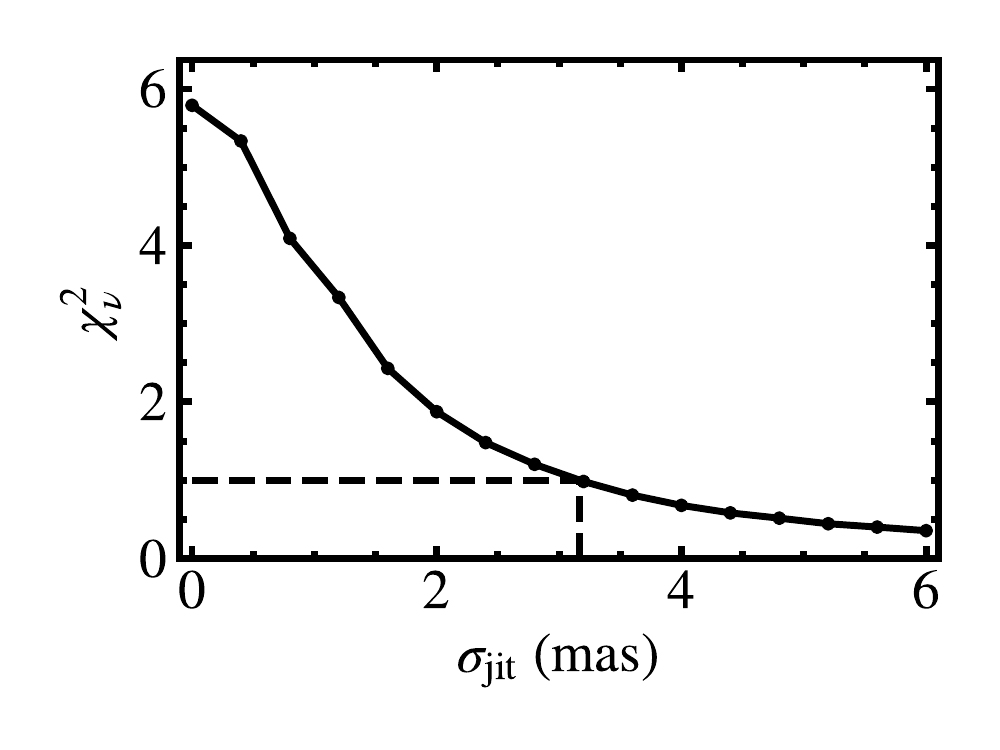}
    \caption{Reduced $\chi^2$ values from orbit fits to the ensemble of relative astrometry for HD 984 as a function of increasing astrometric jitter. An astrometric jitter of $\sigma_{\mathrm{jit}} = \SI{3.17}{mas}$ yields $\chi^2_{\mathrm{\nu}} = 1.0$. We therefore inflate the relative astrometry uncertainties by this amount for the final joint orbit fit.
    \label{fig:chisq_plot}}
\end{figure}

Figure \ref{fig:chisq_plot} shows the result of this procedure. Through 3rd-order spline interpolation, we find that an astrometric jitter value of $\sigma_{\mathrm{jit}} = \SI{3.17}{mas}$ produces $\chi_{\nu}^2 = 1.0$. We adopt this value for the full orbit fit, using Equations \ref{eq:ei_sep} and \ref{eq:ei_pa} to inflate the raw separation and position angle measurements, respectively.

\begin{deluxetable*}{lccc} 
\tablecaption{\label{tab:elements}HD 984 B Orbit Fit Results}
\tablehead{\colhead{Parameter} & \colhead{Median $\pm 1\sigma$} & \colhead{95.4\% C.I.} & \colhead{Prior}}
\startdata
\multicolumn{4}{c}{Fitted Parameters} \\
\hline
$M_{\mathrm{comp}}$ $(\mathrm{M_{Jup}})$ & $61 \pm 4$ & (53, 69) & $1/M_{\mathrm{comp}}$ (log-flat)\\
$M_{\mathrm{host}}$ $(\mathrm{M_\odot})$ & $1.16 \pm 0.10$ & (0.94, 1.36) & $\SI{1.2 \pm 0.1}{M_\odot}$ (Gaussian)\\
$a$ $(\mathrm{AU})$ & ${28}_{-4}^{+7}$ & (21, 46) & $1/a$ (log-flat)\\
$i$ $(\si{\degree})$ & ${120.8}_{-1.6}^{+1.8}$ & (117.6, 124.8) & $\sin (i)$, $\SI{0}{\degree} < i < \SI{180}{\degree}$\\
$\sqrt{e} \sin{\omega}$ & ${-0.62}_{-0.03}^{+0.07}$ & (-0.68, 0.64) & Uniform\\
$\sqrt{e} \cos{\omega}$ & ${0.59}_{-0.08}^{+0.06}$ & (-0.68, 0.72) & Uniform\\
$\Omega$ $(\si{\degree})$ & ${115}_{-7}^{+10}$ & (101, 303) & Uniform\\
$\lambda_{\mathrm{ref}}$ $(\si{\degree})$\tablenotemark{a} & ${312}_{-7}^{+6}$ & (130, 324) & Uniform\\
$\bar{\omega}$ $(\si{mas})$ & $21.877 \pm 0.004$ & (21.868, 21.885) & $\SI{21.877 \pm 0.025}{mas}$ (Gaussian)\\
HPF RV Jitter $\sigma_{\mathrm{HPF}}$ $\mathrm{(m/s)}$ & ${0}_{-0}^{+26}$ & (0, 49) & $1/\sigma_{\mathrm{HPF}}$ (log-flat), $\sigma_{\mathrm{HPF}} \in [0, 1000\si{m/s}]$\\
HARPS RV Jitter $\sigma_{\mathrm{HARPS}}$ $\mathrm{(m/s)}$ & ${89}_{-13}^{+17}$ & (66, 130) & $1/\sigma_{\mathrm{HARPS}}$ (log-flat), $\sigma_{\mathrm{HARPS}} \in [0, 1000\si{m/s}]$\\
\hline
\multicolumn{4}{c}{Derived Parameters} \\
\hline
$P$ (yr) & ${140}_{-30}^{+50}$ & (90, 280) & . . .\\
$e$ & $0.76 \pm 0.05$ & (0.64, 0.85) & . . .\\
$\omega$ $(\si{\degree})$\tablenotemark{b} & ${314}_{-3}^{+4}$ & (308, 323) & . . .\\
$T_0$ $(\mathrm{JD})$ & ${2461600}_{-1700}^{+2500}$ & (2457600, 2468400) & . . .\\
$q$ $(=M_{\mathrm{comp}}/M_{\mathrm{host}})$ & ${0.0500}_{-0.0023}^{+0.0029}$ & (0.0458, 0.0571) & . . .
\enddata
\tablenotetext{a}{Mean longitude at the reference epoch of 2010.0.}
\tablenotetext{b}{Our $\omega$ posterior consists of two distinct peaks separated by $\SI{180}{\degree}$. The values shown in the table correspond to the substantially higher peak; we report summary statistics of this peak by restricting $\omega > \SI{226}{\degree}$. The lower peak is located at ${136}_{-4}^{+5}\si{\degree}$ with a 95.4\% confidence interval of ($\SI{129}{\degree}$, $\SI{147}{\degree}$).}
\end{deluxetable*}

\begin{figure*}
    \centering
    \includegraphics[width=0.65\textwidth]{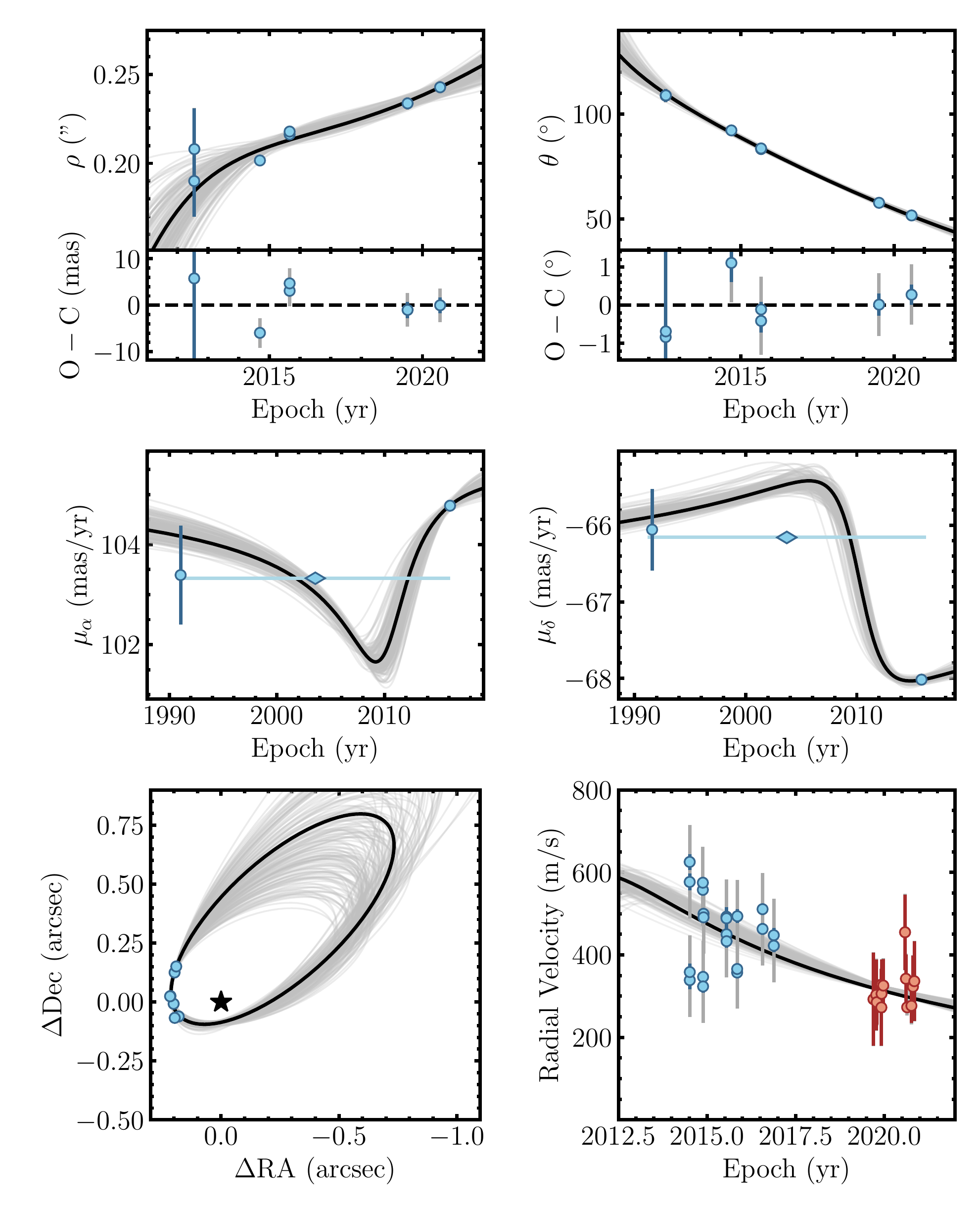}
    \caption{Orbit fit of HD 984 B compared to the relative astrometry (top panels), HGCA acceleration of its host star (middle panels), sky-projected orbit (bottom left), and radial velocities (bottom right). The gray curves show 150 randomly drawn orbits from the MCMC chains. The maximum-likelihood orbit is highlighted in black. In the RV plot, the red points are HPF RVs, while the blue points are HARPS RVs. The gray error bars in the relative astrometry plots are the inflated uncertainties, while the blue error bars are the original uncertainties. The gray error bars in the RV plot incorporate the jitter terms from the best-fit orbit ($\sigma_{\mathrm{HARPS}}$ and $\sigma_{\mathrm{HPF}}$), while the blue and red error bars are the original instrumental uncertainties. \label{fig:orbit_fit}}
\end{figure*}

\begin{figure*}
    \centering
    \includegraphics[width=0.8\textwidth]{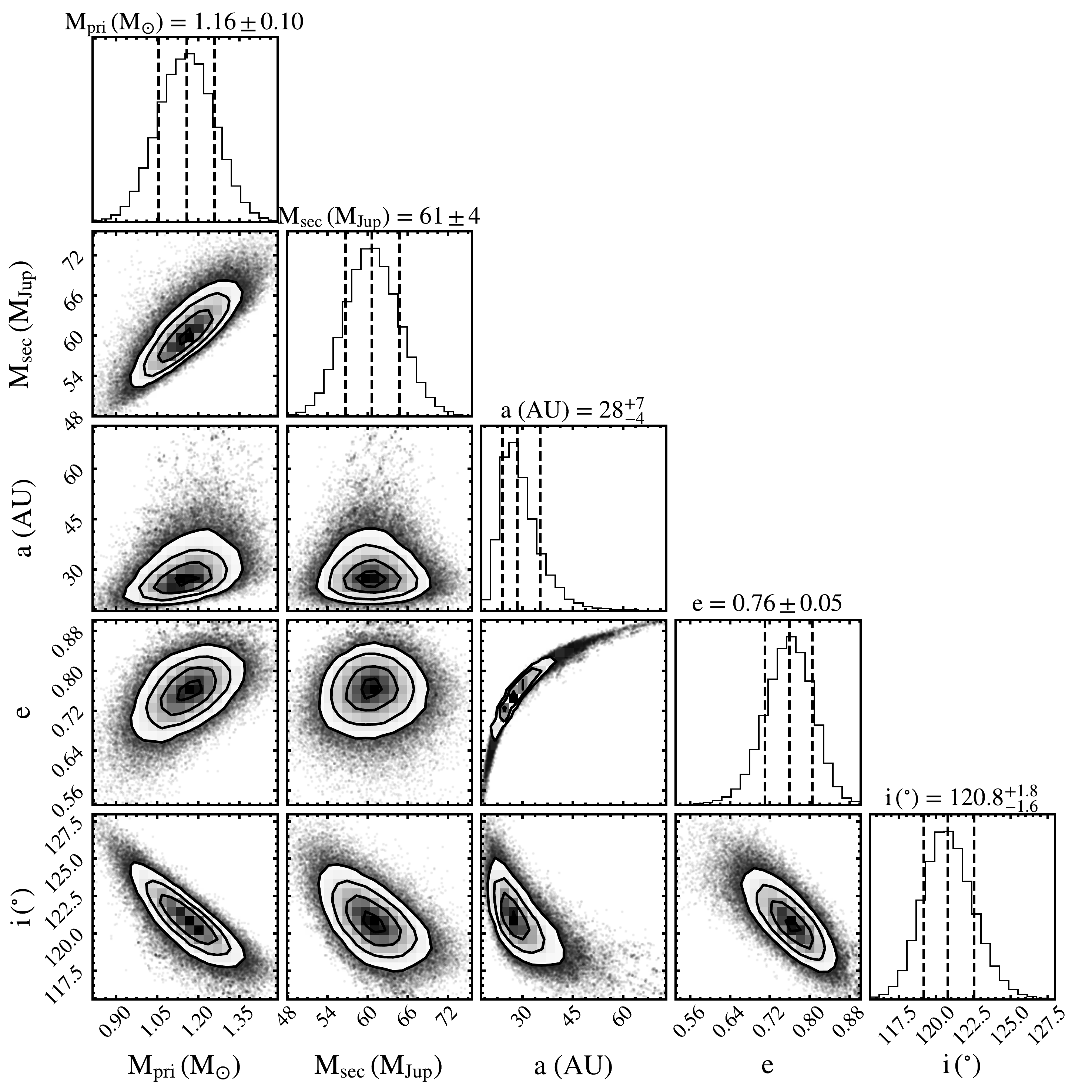}
    \caption{Joint posterior distributions for select parameters in the orbit fit of HD 984 B. Diagonal panels show the marginalized posteriors for each parameter. The dynamical mass of HD 984 B is $61 \pm 4$ $\mathrm{M_{Jup}}$, placing the companion firmly in the brown dwarf regime.
    \label{fig:corner_orbit}}
\end{figure*}

\subsection{3D Orbit}
The mass and orbit of HD 984 B are determined by jointly fitting the relative astrometry, radial velocities, and proper motion difference between \emph{Hipparcos} and \textit{Gaia} using the \texttt{orvara} package\footnote{We used a version of \texttt{orvara} that incorporates Github commits through March 3rd, 2021.} \citep{brandt_orvara_2021}. \texttt{orvara} implements an efficient method of solving Kepler's equation based on the approach described in \citet{raposo-pulido_efficient_2017}. This yields absolute errors below $2\times10^{-14}$ for $e < 0.9999$.

The orbital parameter posteriors are sampled via the parallel-tempering Markov chain Monte Carlo (PT-MCMC) ensemble sampler in \texttt{emcee} \citep{foreman-mackey_emcee_2013}. The following eight quantities define a Keplerian orbit: semimajor axis ($a$),  inclination ($i$), longitude of ascending node ($\Omega$), time of periastron passage ($T_0$), eccentricity ($e$), argument of periastron ($\omega$), and the masses of the two components ($M_{\mathrm{host}}$ and $M_{\mathrm{comp}}$). Eccentricity and argument of periastron are fit as $\sqrt{e}\sin\omega$ and $\sqrt{e}\cos\omega$, which avoids the Lucy-Sweeny bias against circular orbits \citep{lucy_spectroscopic_1971}. \texttt{orvara} fits a single RV jitter term $\sigma_\mathrm{jit}$ to capture RV variations caused by activity that are not reflected in the instrumental uncertainties. The HARPS RVs of HD 984 have a higher scatter than the HPF RVs while also having lower measurement uncertainties, so we modified \texttt{orvara} to fit two RV jitter terms for each instrument, $\sigma_\mathrm{HARPS}$ and $\sigma_\mathrm{HPF}$. The package analytically marginalizes over the barycenter proper motion, the RV zero points for the HARPS and HPF datasets ($\mathrm{RV}_\mathrm{zero}$), and the parallax. Like the \texttt{orbitize!} fits, we used PT-MCMC to fit a total of 20 temperatures, where the coldest set of chains describes the posterior distributions. We found that 100 walkers over $10^6$ steps samples the parameter space well. We discard the first 250,000 steps as burn-in.

Our priors and orbit fit results are listed in Table \ref{tab:elements}. We adopt a prior on the host star mass of $\SI{1.2 \pm 0.1}{M_\odot}$, which encompasses typical mass estimates for HD 984 in the literature (e.g., $\SI{1.2 \pm 0.06}{M_\odot}$ from \citealt{meshkat_discovery_2015}; $\SI{1.15 \pm 0.06}{M_\odot}$ from \citealt{mints_unified_2017}; $\SI{1.28}{M_\odot}$ from \citealt{nielsen_gemini_2019}). For the prior on the parallax, we adopt the \emph{Gaia} EDR3 value of $\SI{21.877 \pm 0.025}{mas}$. All other priors are uninformative and can be found in Table \ref{tab:elements}.

Figure \ref{fig:orbit_fit} overplots the relative astrometry, absolute \emph{Hipparcos}-\emph{Gaia} astrometry, radial velocities, and sky-projected relative offsets on a sample of the orbit solutions. The maximum-likelihood orbit is highlighted in black. Our new relative astrometry in 2019 and 2020 significantly extends the previous orbital arc, resulting in a shallower slope in separation than previous orbit fits that only incorporated the 2012--2015 astrometry (see panel (a) of Figure 13 from \citealt{bowler_population-level_2020}). Note that the middle measurement in the \emph{Hipparcos}-\emph{Gaia} data is an average proper motion based on scaled positions in HGCA and reflects the astrometric reflex motion over the entire 25-year time baseline between the two missions. For clarity, we only show orbits in the RV plot with $\omega > \SI{226}{\degree}$. There are two distinct clusters of orbits with arguments of periastra $\SI{180}{\degree}$ apart that correspond to the star approaching and receding Earth at the time our observations were taken. This is due to the RV datasets lacking the time baseline or RV precision to unambiguously constrain the direction of the radial acceleration.

Joint posteriors of host star mass, companion mass, and select orbital elements are shown in Figure \ref{fig:corner_orbit}. The dynamical mass of HD 984 B is $61 \pm 4$ $\mathrm{M_{Jup}}$, which places it firmly in the brown dwarf regime (${\lesssim} \SI{75}{M_{Jup}}$) at 99.7\% significance. The inclination of $i=120\fdg8^{+1.8}_{-1.6}$ from our fit is consistent with the previous measurement of $i = 120^{+6.1}_{-7.3} {}^{\circ}$ in \citet{bowler_population-level_2020}, while the semi-major axis increases from $a = 17.6^{+4.3}_{-8.1}$ AU from \citet{bowler_population-level_2020} to $a = 28^{+7}_{-4}$ AU. The eccentricity increases substantially between the orbit fits: the previous fit found $e = 0.23^{+0.11}_{-0.23}$, while our new orbit fit yields an eccentricity of $e = 0.76 \pm 0.05$. The difference in $a$ and $e$ can largely be attributed to the increased orbit coverage and astrometric acceleration we make use of in this fit. The higher eccentricity is consistent with the primary conclusion of \citet{bowler_population-level_2020} that brown dwarf companions exhibit a broad eccentricity distribution. This stands in contrast to imaged planets which preferentially have low eccentricities.

Among the relative astrometry datasets, the 2014 SINFONI data largely drives the error inflation. Compared to the other separation measurements, it has the smallest uncertainty by a factor of ${\sim}2$. Excluding the SINFONI astrometry from the \texttt{orbitize!} fit of the isolated relative astrometry produces $\chi^2_{\nu} = 1.1$. To ensure that the inclusion of this point is not significantly influencing the final orbit fit, we performed an additional \texttt{orvara} fit omitting the SINFONI astrometry, which yields $M_{\mathrm{comp}} = 61 \pm 4$ $\mathrm{M_{Jup}}$, $a = 29^{+5}_{-3}$ AU, $e = 0.75 \pm 0.03$, and $i = 119\fdg8 \pm 1\fdg7$. These results are consistent with the orbit fit that includes all astrometry with inflated uncertainties.

The time of periastron for our orbit fit (in calendar years) of $T_0 = {2028}_{-5}^{+7}$ is close to the epoch of our relative astrometry (2012--2020), especially in comparison to the orbital period of $P = {140}_{-30}^{+50}$ yr. From Kepler's second law, eccentric companions spend a significantly smaller fraction of time near periastron compared to apastron. Finding that an eccentric companion happened to be discovered near periastron could be a sign that systematic errors are present in the astrometry. While the companion could of course be in that position in its orbit, another possibility is that this is caused by underestimated errors in the astrometry. Additional systematics not captured in the error inflation would produce larger velocity changes between observations, causing the orbit fit to prefer eccentric solutions with periastron near the epoch of the astrometry. Another explanation is that eccentric companions near periastron produce the most precise orbits over short orbital arcs, causing a selection effect that favors companions with these types of orbits, as they are most amenable to precise dynamical mass measurements.
Further monitoring of this system will determine whether HD 984 B is truly near periastron.
\clearpage
\section{Discussion \label{sec:discussion}}
\subsection{Age of HD 984}
Determining the age of the HD 984 system has proven to be challenging. The spectral type (F7) and effective temperature $T_{\mathrm{eff}} = \SI{6315}{K}$ \citep{casagrande_new_2011} of HD 984 place it at the Kraft break \citep{kraft_studies_1967}, which separates lower-mass stars with deep convective envelopes and higher-mass stars with thin convective envelopes. Convective zones play important roles in generating magnetic winds that drive angular momentum loss. Stars with masses above the Kraft break tend to retain their initial rotation rates, while stars below that threshold spin down over their lifespans. This leads to the breakdown of standard gyrochronological relations such as \citet{mamajek_improved_2008} for stars like HD 984. It also hampers activity-based age indicators, as activity is caused by a star's magnetic dynamo. For instance, \texttt{BAFFLES} \citep{stanford-moore_baffles_2020} returns a broad age posterior with a median value of $\SI{444}{Myr}$ and 95\% confidence interval of 57--6670 Myr for the $\log(R'_{HK}) = -4.34$ dex measurement from \citet{boro_saikia_chromospheric_2018}.

\citet{meshkat_discovery_2015} carried out a detailed age analysis of HD 984 and settled on a range of $30-200$ Myr. The low end is derived from isochronal fitting, which produces a broad age distribution with a lower limit of ${\sim}\SI{30}{Myr}$ set by the time that it takes for $\SI{1.2}{M_\odot}$ stars to settle onto the zero-age main sequence. The upper limit is inferred from HD 984's X-ray luminosity ($\log(L_X/L_{\mathrm{bol}}) = \SI{-4.22}{dex}$; \citealt{voges_rosat_1999, meshkat_discovery_2015}), Ca II H\&K emission (\citealt{meshkat_discovery_2015} adopted a mean value of $\log (R'_{HK})\approx-4.40$ dex), and projected rotational velocity ($v\sin i = 42.13 \pm 1.65$ km/s; \citealt{white_high-dispersion_2007})\footnote{This projected rotational velocity from \citet{white_high-dispersion_2007} is consistent within $2\sigma$ with a recent measurement of $v\sin i = 39.3 \pm 1.5$ km/s from \citet{zuniga-fernandez_search_2021}.}. To overcome the challenges produced by the star's position at the Kraft break, \citet{meshkat_discovery_2015} compare these quantities against a sample of nearby stars with the same spectral type and find that HD 984 is among the youngest $1{-}5$ percent of field F7 dwarfs. This sets an upper limit on the age of HD 984 of ${\sim} 200$ Myr based on the main-sequence lifetime of 1.2 $\mathrm{M_\odot}$ stars of ${\sim} 5000$ Myr.

HD 984 was observed by \emph{TESS} \citep{ricker_transiting_2014} in Sector 3 (TIC 408012706), which presents the possibility of refining its age through gyrochronology or asteroseismology. Its light curve exhibits significant periodicity indicative of rotation. Analyzing the lightcurve with a Lomb-Scargle periodogram and fitting a Gaussian to the highest peak to estimate the uncertainty yields a rotation period of $1.381 \pm 0.012$ d. The \citet{mamajek_improved_2008} relations with $B - V = 0.500 \pm 0.022$ mag \citep{hog_tycho-2_2000} imply an age of $200$ Myr with a $1\sigma$ upper limit of 700 Myr, where the large uncertainty is due to HD 984's position near the Kraft break. Note that the minimum value of $B-V$ for which the \citet{mamajek_improved_2008} relations are valid is $B-V = 0.5$ mag---exactly matching that for HD 984. Asteroseismology offers another possible avenue to determine the age of HD 984. We smoothed and subtracted the \emph{TESS} light curve with a box filter to remove large periodic structure from stellar rotation and examined the residuals. A Lomb-Scargle periodogram of the residuals shows no significant peaks, so the potential to revise the age of HD 984 via asteroseismology is limited with the current precision from \emph{TESS}.

\citet{zuckerman_tucanahorologium_2011} considered HD 984 to be a member of the $42^{+6}_{-4}$ Myr Columba young association primarily on the basis of its $UVW$ kinematics derived from \emph{Hipparcos} astrometry \citep{torres_young_2008, bell_self-consistent_2015}. Newer \emph{Gaia} EDR3 astrometry and the \emph{Gaia} DR2 radial velocity produce $UVW$ space motions of $U = -12.47 \pm 0.03$ km/s, $V = -22.37 \pm 0.19$ km/s, and $W = -8.8 \pm 0.5$ km/s. Compared with the mean velocities of Columba from \citet{gagne_banyan_2018-1} of $(UVW) = (-11.9\pm 1.0, -21.3 \pm 1.3, -5.7 \pm 0.7)$ km/s, HD 984 is discrepant primarily in its $W$ velocity by $3.6\sigma$, or $3.1 \pm 0.9$ km/s. BANYAN $\Sigma$ \citep{gagne_banyan_2018-1} yields a $99.6\%$ probability that HD 984 is a field star based on its \emph{Gaia} EDR3 astrometry and \emph{Gaia} DR2 radial velocity. However, BANYAN $\Sigma$ is susceptible to membership misassignments or omissions as a result of assumptions about cluster spatial and kinematic structure (e.g., \citealt{bowler_planets_2017, desrochers_banyan_2018}), so HD 984 could potentially still be a member of the association. 
We used \texttt{FriendFinder}\footnote{https://github.com/adamkraus/Comove} to search for potential co-moving neighbors (either distantly bound or as a cluster) of HD 984 that could help age-date the system (e.g., \citealt{tofflemire_tess_2021}). This tool queries the \emph{Gaia} EDR3 source catalog for apparent comoving stars within a $\SI{25}{pc}$ volume that have a tangential (sky-plane) velocity offset from HD 984 B's proper motion of $ {<} \SI{5}{km/s}$. No overdensity of stars above the main sequence consistent with membership in a young moving group is apparent. Based on the lack of a convincing kinematic match to Columba or other young comoving stars, we adopt the age estimate of 30--200 Myr from \citet{meshkat_discovery_2015} for HD 984 A and B.

\subsection{Bolometric Luminosity of HD 984 B}
We calculate the bolometric luminosity of HD 984 B via the $M_{H,\mathrm{2MASS}}-\log(L_{\mathrm{bol}} / L_\odot)$ relation from \citet{dupuy_individual_2017}. We first convert the $H$-band magnitude of the companion from the MKO photometric system to the 2MASS photometric system using the empirical $H_{\mathrm{MKO}} - H_{\mathrm{2MASS}}$ formula from \citet{dupuy_individual_2017}. Converting to an absolute magnitude using the \emph{Gaia} EDR3 parallax yields $M_{H,2MASS} = 9.27 \pm 0.05$ mag, which corresponds to a bolometric luminosity of $\log(L_{\mathrm{bol}}/L_\odot) = -2.88 \pm 0.03$ dex. Although the \citet{dupuy_individual_2017} luminosity relation is based on dynamical masses of field objects, bolometric corrections for field and young ultracool dwarfs agree for mid- and late-M dwarfs \citep{filippazzo_fundamental_2015}.  Our value agrees with both the luminosity that \citet{johnson-groh_integral_2017} determined by fitting DUSTY models \citep{chabrier_evolutionary_2000} to $J$- and $H$-band photometry of the companion ($-2.88 \pm 0.07$ dex) and the value from \citet{meshkat_discovery_2015} who applied M6.5 dwarf bolometric corrections to NaCo and SINFONI photometry ($-2.815 \pm 0.024$ dex). We adopt our revised value for this analysis.

\begin{figure}
    \centering
    \includegraphics[width=0.95\linewidth]{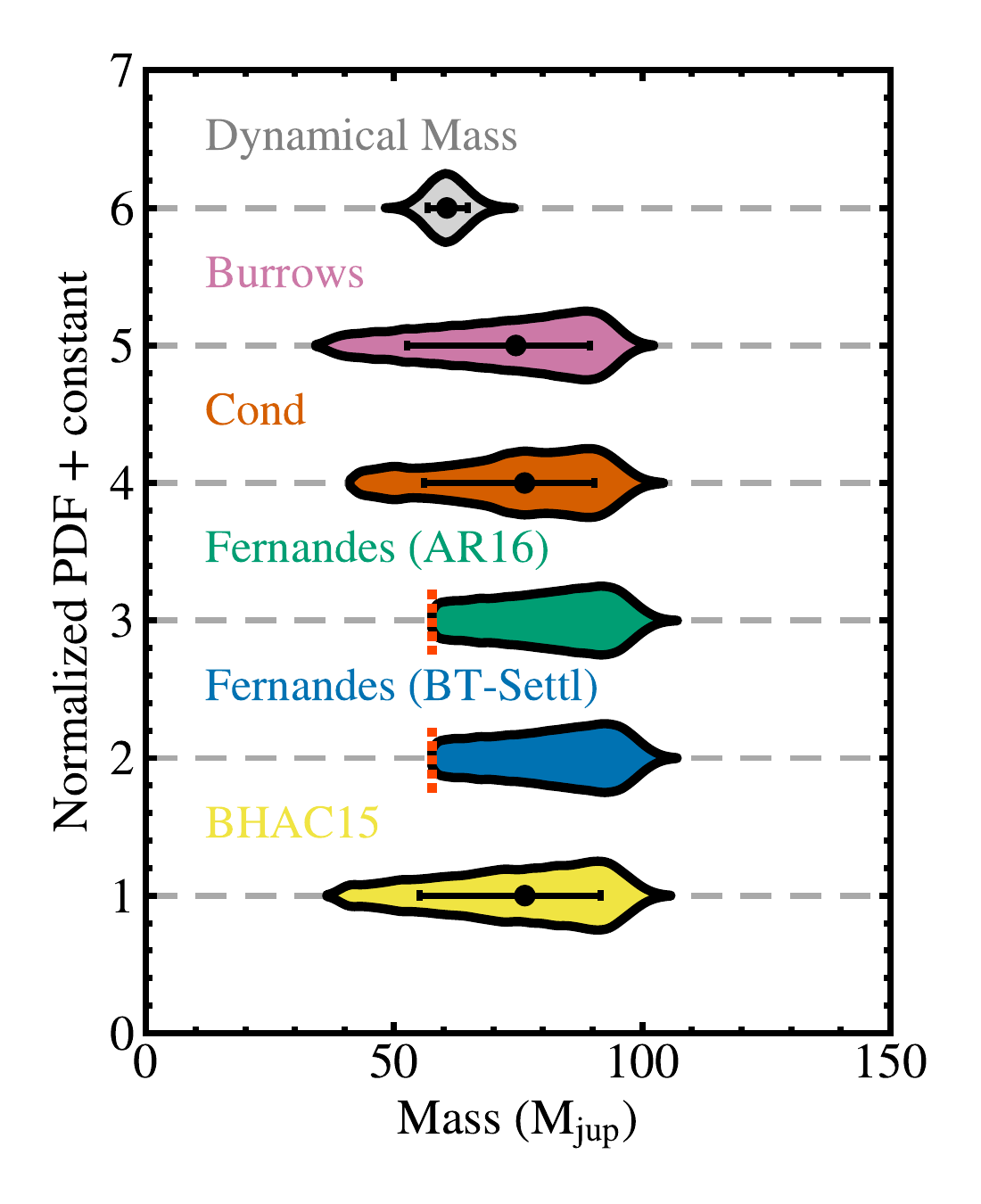}
    \caption{Dynamical mass of HD 984 B (top distribution) compared to inferred mass distributions for a variety of hot-start evolutionary models. Mass distributions are assembled for a given model by sampling the luminosity and age distributions of HD 984 B $10^6$ times, and then linearly interpolating the model grid to produce a corresponding mass for each trial. For the \citet{fernandes_evolutionary_2019} models, the predicted mass distributions are limited to ${>} \SI{57.6}{M_{Jup}}$, which is the lowest-mass track computed in their study (see Figure \ref{fig:tracks}). These inferred masses are thus overestimates of the true model predictions for these two grids. The median and 68.3\% confidence interval are shown for each complete distribution. Each model prediction is broadly consistent with our measured dynamical mass.\label{fig:model_comp}}
\end{figure}

\subsection{Model Comparison}
Our dynamical mass measurement of HD 984 B can be directly compared with the predicted masses of substellar evolutionary models. A variety of publicly available hot-start\footnote{We only consider hot-start models for this analysis, as cold- and warm-start grids generally do not extend to HD 984 B's luminosity and age.} evolutionary models are selected for this comparison based on the luminosity and age of HD 984 B. Specifically, we generate inferred mass distributions from the pre-computed grids of \citet{burrows_nongray_1997}, \citet{baraffe_evolutionary_2003} (Cond\footnote{We do not consider DUSTY \citep{chabrier_evolutionary_2000}, the dust opacity-dominated version of Cond, since grain opacity is insignificant for objects earlier than ${\sim}\mathrm{M8}$ or $T_{\mathrm{eff}} \gtrsim 2570$ \citep{allard_limiting_2001}.}), \citet{baraffe_new_2015} (BHAC15), and \citet{fernandes_evolutionary_2019}. Note that the \citet{fernandes_evolutionary_2019} grids only partially cover the luminosity and age of HD 984 (see Figure \ref{fig:tracks}).

In general, evolutionary models are constructed by coupling an interior structure model to an atmospheric model as a surface boundary condition (see \citealt{marley_cool_2015} for a review on radiative-convective substellar atmospheric models). The hot-start models we consider here vary in their input physics, equations of state, compositions, and atmospheric models. The \citet{burrows_nongray_1997} and Cond grids both use the same H/He equation of state from \citet{saumon_equation_1995}. Differences in internal structure between the models originate from the assumed helium fractions (which impacts deuterium and lithium burning) and initial conditions (for ${\lesssim}\SI{1}{Gyr}$ ages). In addition, the atmospheric models utilized in these evolutionary codes differ. \citet{burrows_nongray_1997} use gray atmospheres from \citet{saumon_theory_1996} for $T_{\mathrm{eff}} \gtrsim \SI{1300}{K}$ (which includes HD 984 B) and non-gray model atmospheres computed in a manner similar to \citet{marley_atmospheric_1996} for lower $T_{\mathrm{eff}}$. Cond computes non-gray atmospheres through the \texttt{PHOENIX} code \citep{allard_model_1995, hauschildt_nextgen_1999}, making modifications to incorporate condensation for T dwarfs. These differences in atmospheric modeling have a small impact on the resulting luminosity evolution, however, as luminosity is a shallow function of the Rosseland mean opacity ($L\sim \kappa_\mathrm{R}^{0.35}$; \citealt{burrows_science_1993}), so changes in the opacity from different treatments of clouds have a relatively weak effect on the thermal evolution.

The BHAC15 grid of \citet{baraffe_new_2015} uses the same interior structure model as Cond, but with an updated treatment of convection and atmospheric boundary condition \citep[BT-Settl;][]{allard_models_2012,  allard_atmospheres_2012}. The \citet{fernandes_evolutionary_2019} models adapt the stellar evolution code CLES \citep{scuflaire_cles_2008} for ultracool dwarfs, adopting either the BT-Settl models or model atmospheres from \citet{aringer_synthetic_2016}. The primary difference between the two atmospheric models is that BT-Settl (calculated using a newer version of the \texttt{PHOENIX} code) includes grain formation for $T_{\mathrm{eff}} < \SI{2600}{K}$, while the \citet{aringer_synthetic_2016} grid is dust free. Other, more recently released hot-start substellar model grids generally do not reach the high effective temperature and luminosity of HD 984 B at its young age \citep[e.g.,][]{saumon_evolution_2008, phillips_new_2020}.

\begin{figure}
    \centering
    \includegraphics[width=0.92\linewidth]{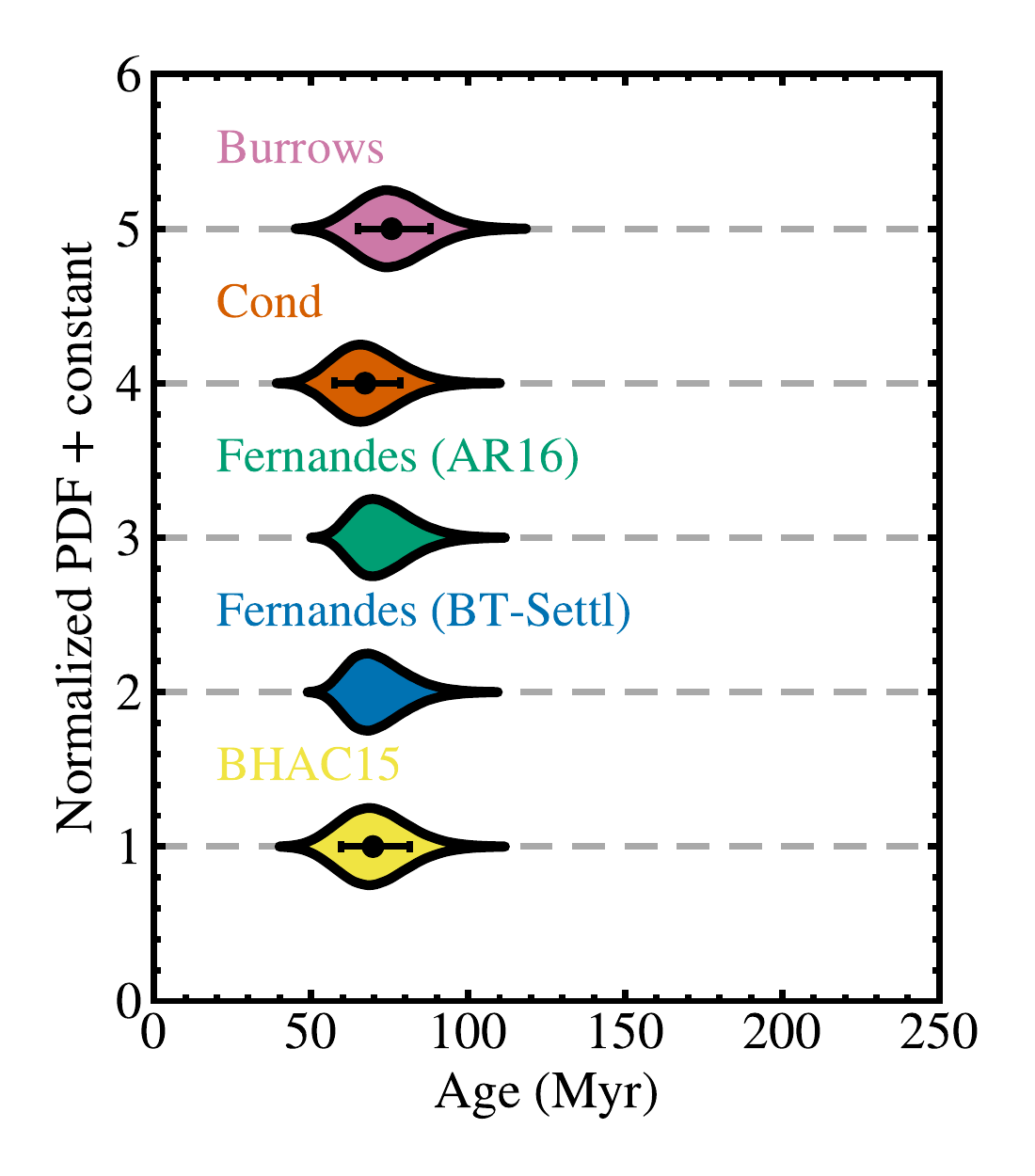}
    \caption{Predicted ages of HD 984 B based on the dynamical mass and luminosity for a variety of hot-start evolutionary models. Ages are inferred for a given model by sampling the luminosity and dynamical mass distributions of HD 984 B $10^6$ times, and then linearly interpolating the model grid to produce a corresponding age for each trial. Due to the partial coverage of the \citet{fernandes_evolutionary_2019} grids over HD 984 B's luminosity and mass (see Figure \ref{fig:tracks}), the corresponding predicted ages are somewhat overestimated for these models. Note that there is not a strict minimum age for these two models because the inferred age is luminosity-dependent. Altogether, hot-start models predict a young age of ${\sim}50{-}90$ Myr.\label{fig:model_age}}
\end{figure}

\begin{figure*}
    \centering
    \includegraphics[width=0.85\textwidth]{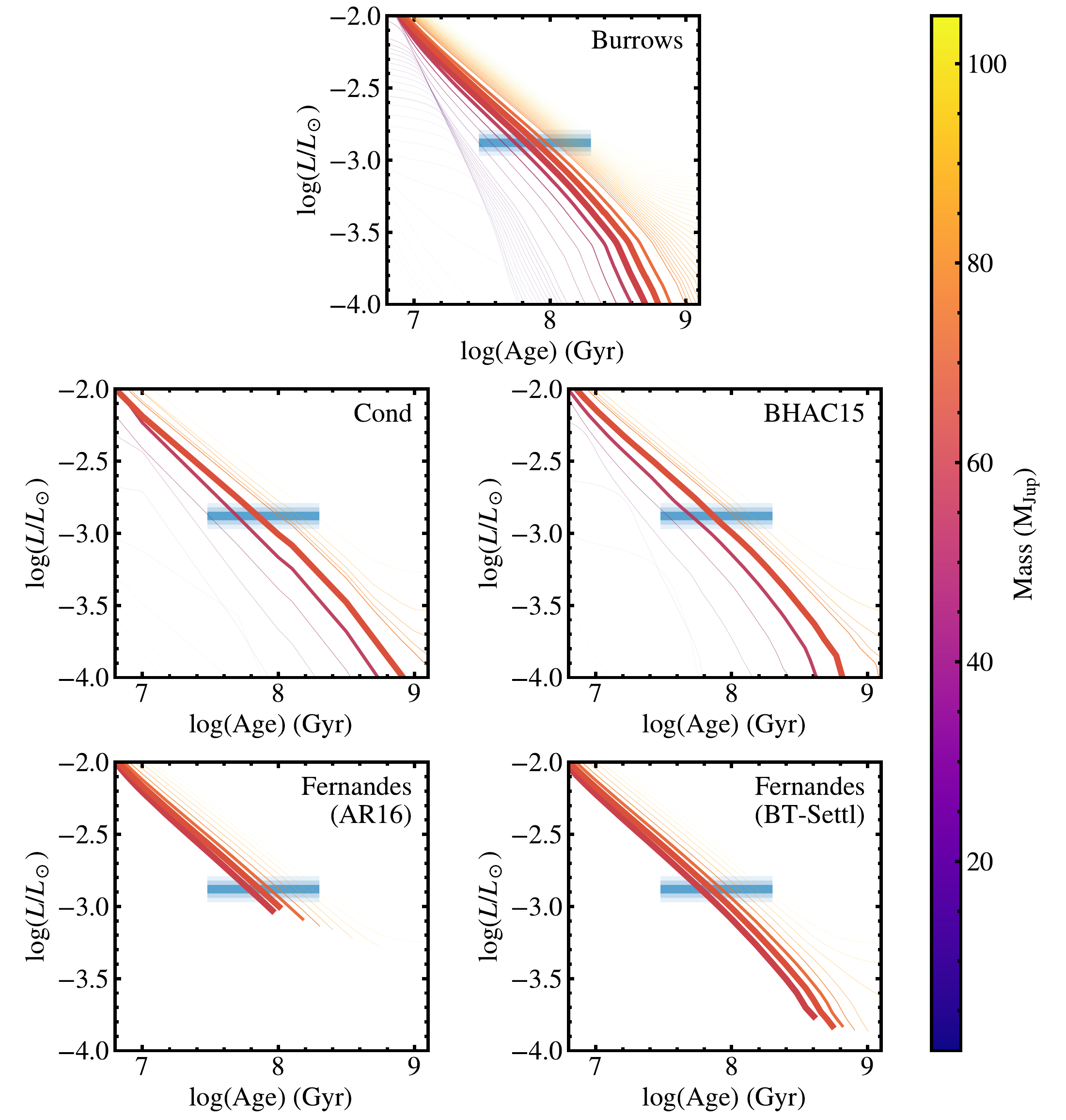}
    \caption{Iso-mass tracks from each evolutionary model considered in this study compared with HD 984 B's age and luminosity. We adopt a linearly uniform distribution from 30--200 Myr for HD 984 B's age and a Gaussian distribution for its luminosity. Contours of 1, 2, and $3\sigma$ are shown in blue. The color of the model tracks corresponds to their mass. The transparency and thickness of the tracks is related to the dynamical mass posterior probability density. The narrow dynamical mass distribution ($61 \pm 4$ $\mathrm{M_{Jup}}$) means that only a few model tracks available from these grids appear thick in each panel. Note that the \citet{fernandes_evolutionary_2019} models only partially cover the luminosity, age, and dynamical mass of HD 984 B.\label{fig:tracks}}
\end{figure*}

We employ a Monte Carlo approach to generate predicted mass distributions for each evolutionary model grid. The luminosity and age distributions of HD 984 B are first sampled $10^6$ times. The model grids are then linearly interpolated to produce a mass for each luminosity and age trial. We do not extrapolate outside of the model grids; this is pertinent to the \citet{fernandes_evolutionary_2019} models, which only cover a portion of HD 984 B's luminosity and age range (see Figure \ref{fig:tracks}). Solar metallicity grids are adopted for a uniform comparison of the models and because HD 984 is only slightly metal-rich ($\mathrm{[Fe/H]} = 0.27$; \citealt{luck_abundances_2018}). In the case of \citet{fernandes_evolutionary_2019}, we compute predicted mass distributions for both BT-Settl and \citet{aringer_synthetic_2016} model atmospheres. 

Figure \ref{fig:model_comp} shows the resulting inferred mass distributions for HD 984 B. The broad age constraint results in correspondingly broad predicted mass distributions for both models. Our dynamical mass measurement is consistent within $1\sigma$ of all model predictions. Note that the \citet{fernandes_evolutionary_2019} predicted mass distributions are limited to ${>} \SI{57.6}{M_{Jup}}$, which corresponds to the lowest mass track computed for those models. The Cond and BHAC15 inferred masses are similar, owing to their use of nearly identical interior structure models. The two atmospheric treatments with the \citet{fernandes_evolutionary_2019} grids also produce similar inferred masses. Overall, the broad age constraint for HD 984 limits detailed comparisons with evolutionary models. Improved age estimates would yield more precise inferred mass distributions and enable more rigorous model tests.

We also consider what ages the various models predict for HD 984 B given its dynamical mass and luminosity. Figure \ref{fig:model_age} shows the result of sampling the luminosity and dynamical mass distributions $10^6$ times through linear interpolation to find the corresponding age for each trial and model. Each model predicts a young age for HD 984 B of ${\sim}50{-}90$ Myr. This is consistent with the age constraint for HD 984 of $30{-}200$ Myr. If the true age of HD 984 is near the upper bound of that range, the companion could be overluminous, as was found for Gl 417 BC \citep{dupuy_new_2014} and HD 130948 BC \citep{dupuy_dynamical_2009}. Note again that the \citet{fernandes_evolutionary_2019} grids only partially cover the properties of HD 984 B, so the age distributions somewhat overestimate the predicted age for these two models. This is demonstrated in Figure \ref{fig:tracks}, which displays the iso-mass tracks of each model against HD 984 B's luminosity and age constraint. 
% \clearpage
\subsection{Mass of HD 984 B}
Our mass measurement of $61 \pm 4$ $\mathrm{M_{Jup}}$ places HD 984 B well within the brown dwarf regime. The precise hydrogen burning limit (HBL), defined as the point where half of an object's luminosity is generated via hydrogen fusion \citep{reid_new_2005}, is a function of opacity (and by extension metallicity) and helium fraction \citep{burrows_theory_2001}.\footnote{Rotation rate can also influence the HBL, with slower rotating objects able to sustain hydrogen burning at lower masses \citep{chowdhury_stable_2021}.} Stars with higher opacities, metallicities, and helium fractions can sustain hydrogen burning at lower masses. For example, the CLES models \citep{scuflaire_cles_2008} upon which the \citet{fernandes_evolutionary_2019} models are based predict hydrogen burning mass limits that range from $\SI{78}{M_{Jup}}$ for $\mathrm{[Fe/H]}=+0.5$ to $\SI{84}{M_{Jup}}$ for $\mathrm{[Fe/H]}=-0.3$. The hydrogen burning limit can vary among models as well. For the CLES models, the substellar boundary is ${\sim} \SI{82}{M_{Jup}}$ for solar metallicity, while the BHAC15 models \citep{baraffe_new_2015} produce a hydrogen burning limit of ${\sim}\SI{73}{M_{Jup}}$ and the \citet{saumon_evolution_2008} models produce a limit of ${\sim}{\SI{70}{M_{Jup}}}$. \citet{dupuy_individual_2017} empirically constrained the substellar boundary to be $70 \pm 4$ $\mathrm{M_{Jup}}$ by measuring dynamical masses of ultracool dwarf binaries. At a fiducial value of $75$ $\mathrm{M_{Jup}}$ for the HBL, HD 984 B's dynamical mass places it in the brown dwarf regime to 99.7\% significance. If instead we adopt a lower value of $\SI{70}{M_{Jup}}$ for the HBL, the significance drops slightly to 98.2\%.

Among the small collection of substellar dynamical masses with well-constrained luminosities and ages (see Figure \ref{fig:dm_lum_age} and Table \ref{tab:dm}), HD 984 B is a young analog of several well-studied companions. HD 4113 C is a late T-dwarf benchmark companion with a similar measured mass of $66^{+5}_{-4}$ $\mathrm{M_{Jup}}$ \citep{cheetham_direct_2018}. However, its dynamical mass is discrepant with the hot-start prediction of $36\pm 5$ $\mathrm{M_{Jup}}$, inferred via Cond models with a system age of $5^{+1.3}_{-1.7}$ Gyr. One possible explanation of this would be that HD 4113 C is an unresolved brown dwarf binary. Spectral fitting of this companion produces a radius of ${\sim}\SI{1.5}{R_{Jup}}$, which is significantly higher than expected for its effective temperature and age. The brown dwarf binary scenario could potentially resolve this tension as well, as it would produce higher radius estimates without significantly modifying the shape of the spectrum. 

Another benchmark with a similar mass to HD 984 B but an older age is HD 4747 B. Discovered by \citet{crepp_trends_2016}, this companion is near the L/T transition (T$1\pm2$; \citealt{crepp_gpi_2018}) with an age just under 4 Gyr. Its dynamical mass of $66.3^{+2.5}_{-3.0}$ $\mathrm{M_{Jup}}$ \citep{brandt_precise_2019} agrees well with hot-start model predictions with the exception of \citet{burrows_nongray_1997}. Other model-independent masses near the substellar boundary include Gl 229 B ($71.4\pm 0.6$ $\mathrm{M_{Jup}}$; \citealt{nakajima_discovery_1995, brandt_improved_2021}), HD 72946 B ($72.5 \pm 1.3$ $\mathrm{M_{Jup}}$; \citealt{bouchy_sophie_2016, brandt_improved_2021}), HR 7672 B ($72.7\pm 0.8$ $\mathrm{M_{Jup}}$; \citealt{liu_crossing_2002, crepp_dynamical_2012, brandt_precise_2019}), and HD 47127 B ($105^{+18}_{-27}$ $\mathrm{M_{Jup}}$; \citealt{bowler_mcdonald_2021-1}). As additional dynamical masses are measured for objects with similar masses, the full evolution of brown dwarfs like HD 984 B will be progressively better constrained.

\section{Conclusions \label{sec:conclusion}}
In this work, we measured the dynamical mass of the imaged substellar companion HD 984 B. Previous model-inferred masses that were based on the age and luminosity of the system ranged from 34 to 94 $\mathrm{M_{Jup}}$, which encompasses both the brown dwarf and stellar mass regimes. Our model-independent measurement of $61 \pm 4$ $\mathrm{M_{Jup}}$ places the companion's mass firmly in the brown dwarf regime (${\lesssim} 75$ $\mathrm{M_{Jup}}$) to 99.7\% significance. This dynamical mass was obtained by combining HD 984's astrometric acceleration in the HGCA, new Keck/NIRC2 AO imaging of the companion, and RVs from HPF and HARPS. HD 984 B joins a small but growing list of benchmark substellar companions with measured dynamical masses and well-constrained ages and luminosities. Additional companion mass measurements spanning large regions of mass, age, and luminosity will enable holistic tests of substellar evolutionary models, ultimately improving our understanding of the formation and thermal evolution of brown dwarfs and giant planets.

\begin{acknowledgements} 

K.F. acknowledges support from the National Science Foundation Graduate Research Fellowship Program under Grant No. DGE-1610403. B.P.B. acknowledges support from the National Science Foundation grant AST-1909209 and NASA Exoplanet Research Program grant 20-XRP20$\_$2-0119.

This work was supported by a NASA Keck PI Data Award, administered by the NASA Exoplanet Science Institute. Data presented herein were obtained at the W. M. Keck Observatory from telescope time allocated to the National Aeronautics and Space Administration through the agency's scientific partnership with the California Institute of Technology and the University of California. The Observatory was made possible by the generous financial support of the W. M. Keck Foundation.

These results are based on observations obtained with the Habitable–Zone Planet Finder Spectrograph on the 9.2-meter Hobby-Eberly Telescope at McDonald Observatory. The authors thank the resident astronomers and telescope operators at the HET for obtaining these observations. The Hobby-Eberly Telescope (HET) is a joint project of the University of Texas at Austin, the Pennsylvania State University, Ludwig-Maximilians-Universität München, and Georg-August-Universität Göttingen. The HET is named in honor of its principal benefactors, William P. Hobby and Robert E. Eberly.

The authors wish to recognize and acknowledge the very significant cultural role and reverence that the summit of Maunakea has always had within the indigenous Hawaiian community. We are most fortunate to have the opportunity to conduct observations from this mountain.
\end{acknowledgements}

\facilities{Keck:II (NIRC2), HET (HPF)}
\software{\texttt{VIP} \citep{gomez_gonzalez_vip:_2017}, \texttt{orvara} \citep{brandt_orvara_2021}, \texttt{orbitize!} \citep{blunt_orbitize_2020}, \texttt{ccdproc} \citep{craig_astropyccdproc_2017}, \texttt{photutils} \citep{bradley_astropyphotutils_2019}, \texttt{skimage} \citep{van_der_walt_scikit-image_2014}, \texttt{astropy} \citep{astropy_collaboration_astropy_2013, astropy_collaboration_astropy_2018}, \texttt{pandas} \citep{mckinney_data_2010}, \texttt{matplotlib} \citep{hunter_matplotlib_2007}, \texttt{numpy} \citep{harris_array_2020}, \texttt{scipy} \citep{virtanen_scipy_2020}, \texttt{emcee} \citep{foreman-mackey_emcee_2013}, \texttt{corner} \citep{foreman-mackey_cornerpy_2016}, \texttt{Lightkurve} \citep{lightkurve_collaboration_lightkurve_2018}}

\bibliography{references_new}{}
\bibliographystyle{aasjournal}

\end{document}